\documentclass[prd,reprint,superscriptaddress]{revtex4-1}
\usepackage{amsmath}
\usepackage{amssymb}
\usepackage{graphicx}
\usepackage[caption=false]{subfig}
\usepackage{color}
\usepackage[percent]{overpic}
\usepackage{bm}
\usepackage{rotating}
\usepackage{hyperref}
\usepackage{pbox}
\usepackage{array}
\usepackage{physics}
\usepackage{mathtools}
\usepackage{enumerate}
\usepackage{leftidx}

\usepackage[normalem]{ulem}

\newcommand{\ii}{\mathrm{i}}

\usepackage{suffix}
\DeclarePairedDelimiterX\MeijerM[3]{\lparen}{\rparen}%
{\begin{smallmatrix}#1 \\ #2\end{smallmatrix}\delimsize\vert\,#3}

\newcommand\MeijerG[8][]{%
	G^{\,#2,#3}_{#4,#5}\MeijerM[#1]{#6}{#7}{#8}}

\WithSuffix\newcommand\MeijerG*[7]{%
	G^{\,#1,#2}_{#3,#4}\MeijerM*{#5}{#6}{#7}}

\DeclareMathOperator*{\SumInt}{%
\mathchoice%
  {\ooalign{$\displaystyle\sum$\cr\hidewidth$\displaystyle\int$\hidewidth\cr}}
  {\ooalign{\raisebox{.14\height}{\scalebox{.7}{$\textstyle\sum$}}\cr\hidewidth$\textstyle\int$\hidewidth\cr}}
  {\ooalign{\raisebox{.2\height}{\scalebox{.6}{$\scriptstyle\sum$}}\cr$\scriptstyle\int$\cr}}
  {\ooalign{\raisebox{.2\height}{\scalebox{.6}{$\scriptstyle\sum$}}\cr$\scriptstyle\int$\cr}}
}

\allowdisplaybreaks

\usepackage[export]{adjustbox}

\begin{document}
	
	\title{Quantum delocalization, gauge and quantum optics:\\ The light-matter interaction in relativistic quantum information}

	\author{Richard Lopp}
	\affiliation{Department of Applied Mathematics, University of Waterloo, Waterloo, Ontario, N2L 3G1, Canada}
	\affiliation{Institute for Quantum Computing, University of Waterloo, Waterloo, Ontario, N2L 3G1, Canada}

	\author{Eduardo Mart\'{i}n-Mart\'{i}nez}
	\affiliation{Department of Applied Mathematics, University of Waterloo, Waterloo, Ontario, N2L 3G1, Canada}
	\affiliation{Institute for Quantum Computing, University of Waterloo, Waterloo, Ontario, N2L 3G1, Canada}
	\affiliation{Perimeter Institute for Theoretical Physics, 31 Caroline St N, Waterloo, Ontario, N2L 2Y5, Canada}

\begin{abstract}
We revisit the interaction of a first-quantized atomic system (consisting of two charged quantum particles) with the quantum electromagnetic field, pointing out the subtleties related to the gauge nature of electromagnetism and the effect of multipole approximations. We connect the full minimal-coupling model with the typical effective models used in quantum optics and relativistic quantum information such as the Unruh-DeWitt (UDW) model and the dipole coupling approximation. We point out in what regimes different degrees of approximation are reasonable and in what cases effective models need to be refined to capture the features of the light-matter interaction. This is particularly important when considering the center of mass (COM) of the atom as a quantum system that can be delocalized over multiple trajectories. For example, we show that the simplest UDW approximation with a quantum COM fails to capture crucial R\"ontgen terms coupling COM and internal atomic degrees of freedom with each other and the field. Finally we show how effective dipole interaction models can be covariantly prescribed for relativistically moving atoms.
\end{abstract}
	
	\maketitle
\section{Introduction}
The interaction of matter with light presents two important challenges when trying to find simple models to describe it: the relativistic, covariant, vector nature of light, and the fact that electromagnetism is a gauge theory. Regarding the relativistic nature of the theory, in atomic physics and quantum optics, matter is usually treated non-relativsitically (atoms are, to a good approximation, systems of bound nuclei and low-energy electrons), and thus for simplicity one combines in the same model a relativistic field interacting with a non-relativistic atom.

The gauge dependence of the theory is trickier. It has been a source of issues in simple models of light-matter interaction. Directly using minimal coupling $\hat{\bm p}\cdot\hat{\bm A}$ between charged particles and the EM field together with gauge independent atomic wavefunctions leads to nonphysical, gauge-dependent atomic transition probabilities \cite{lamb,Scully,funai}. These issues have been the subject of a great deal of studies and can be partially overcome by recasting the interaction in terms of a  multipolar Hamiltonian. This is achieved through combinations of canonical and gauge transformations in order to express the interaction in terms of well-known textbook charge-in-a-Coulomb-potential terms and the observable fields $\hat{\bm E}$ and $\hat{\bm B}$ rather than $\hat{\bm A}$. This can be done for  external classical fields with the Goeppert-Mayer transformation \cite{Goeppert}, as well as for quantized electromagnetic (radiation) fields \cite{Babiker83,Babiker84,Babiker93, woolley80}. In the quantum electromagnetic case, the class of transformations employed to arrive at a multipolar Hamiltonian is known as Power-Zienau-Wolley (PZW) transformations. This is the origin of the ubiquitously used `dipole approximation' $\hat{\bm d}\cdot\hat{ \bm E}$. However, there are a number of subtleties to deal with before arriving at this simple dipole coupling Hamiltonian. These subtleties can be relevant in quantum optics, and particularly so in the context of relativistic quantum information (RQI) when we model the interaction of a microscopic, moving atomic probe with the electromagnetic field. In those cases, the multipolar Hamiltonian with quantized fields, even in the dipole approximation, contains the so-called R\"ontgen term which couples the center-of-mass (COM) degrees of freedom of the atom with its internal degrees of freedom and the electromagnetic field, and that is not commonly considered in RQI studies. However, if one wants to model atomic physics, this kind of terms can only be neglected in a few select scenarios.

Indeed, in \cite{Babiker93} it was argued that the R\"ontgen term is required for energy-momentum conservation and gauge invariance of radiation-induced mechanical forces. This is a consequence of the mechanical momentum not coinciding with the canonical momentum of the COM position for ions.
It has been shown, further, that, for classical~\cite{Wilkens93} and quantum~\cite{Wilkens94} COM degrees of freedom, the R\"ontgen term is already necessary to  leading order in the velocity, $v/c$, so as to avoid nonphysical atomic-velocity dependence in the angular distribution of spontaneously emitted photons. In~\cite{BBB2002} and~\cite{Barnett2003} it was then shown that the total spontaneous emission rate (as given by Fermi's Golden rule) for a classical COM under uniform motion requires the inclusion of the R\"ontgen term. Features of this R\"ontgen term have also been explored in~\cite{Barnett2017} for classical fields and classical COM degrees of freedom. The contribution is usually smaller than radiation-pressure forces, but is nonetheless required for correct physical results. In~\cite{Barnett2017friction} it was shown that for a quantum COM the time derivative of the expectation value of the canonical momentum of the COM is observer dependent, at odds with the necessary covariance of predictions. The resolution was found in the inclusion of the atomic binding energy terms in the Hamiltonian. As \cite{Barnett2017friction} noted, the coupling of COM degrees of freedom and a dynamical mass-energy term is a feature missing from the multipolar Hamiltonian. Sonnleitner and Barnett go on in~\cite{barnett2018} to include a low-order relativistic correction for the multipolar Hamiltonian which remedies the absence of the missing dynamical mass-energy.

In the regimes commonly analyzed in relativistic quantum information finite-time couplings can excite the atom out of its interaction with the vacuum (See, among many others, \cite{exc1,Loukosatz,pad96}). This is important, because even though the dipole approximation is a rather common one, it is usually obtained claiming the existence of some characteristic wavelength that dominates the process. The reasoning is then that if the atom is small enough as compared to the dominant wavelength, we can approximate it by a pointlike object and take only the first term on a multipole expansion, something that is not possible to justify when studying vacuum fluctuations.  In those contexts, it was argued in \cite{funai} that a multipole (and in particular a dipole) approximation can indeed be justified if the duration of the interaction is much larger than the light-crossing time of the atom. In a few words,  the frequencies that take part in a vacuum excitation process are suppressed with the tails of the Fourier transforms of the functions encoding the time-dependence of the coupling, as well as the spatial smearing of the atom. It was then shown in~\cite{funai} that if the interaction times are much longer than the characteristic length of the atom's wavefunction, the suppression of the shorter frequencies is strong enough for a dipole expansion to be a good approximation.

A number of subtleties in the multipole (including dipole) approximation appear when carefully considering the role of gauge transformations in the light-matter interaction, and the fact that atoms can actually have a spatial extension since they are not pointlike objects (even in the dipole approximation). Although there has been a plethora of previous work on multipole approximations (above all considering either classical EM fields e.g. \cite{Barnett2017} and/or semiclassical atoms e.g. \cite{Wilkens93,BBB2002,Barnett2003}, with only a few fully quantum setups, e.g.,~\cite{Barnett2017friction}), the  considerations of gauge issues, finite size of the atomic wavefunction (even for dipoles) and possible quantum delocalization of the center of mass are not commonly combined in any previous work known to the authors. The few works that consider a more complete approach regarding gauge and the quantum nature of the interaction (e.g., \cite{Barnett2017friction}) do particularize to eigenstates of the COM and also consider the rotating-wave approximation, which is incompatible with most RQI setups~\cite{funairwa,causality}.  Within the context of RQI, gauge and COM dynamics considerations are not usually present in most of the traditional light-matter interaction models, making it useful to contextualize the particle detector models used in RQI with a complete description of the light-matter interaction.

In this work, we wish to analyze effective models that can capture realistic dynamics of a first-quantized atom interacting with the quantum EM field. This includes a quantized COM, the quantum nature of the atomic multipole operator, and not assuming either the single-mode or rotating-wave approximation, nor taking a discrete field-momentum spectrum in free space. We will take into account recent results by Stritzelberger and Kempf~\cite{stritzelberger} (followed up on in~\cite{nadine2}) where they studied precisely the influence on the atomic dynamics of the initial delocalization of the COM. We will extend those studies to show the extra considerations that one needs in order for the predictions of the model to be gauge-independent and to include the effect of R\"oentgen terms. As a particular example, we will illustrate the effect of the R\"ontgen term in atomic transition rates in the presence of initial COM delocalization.

In particular, we will show that there is only one scenario where one can neglect the R\"ontgen term: when one considers the atomic COM degrees of freedom to be classical, the atoms are tightly localized, and there exists a common rest frame for all the moving atoms in which the R\"ontgen term vanishes. This is for example the case of entanglement harvesting for comoving inertial atoms (see, e.g., \cite{pozas2}), or a single atom when we work in the detector's COM frame for not very relativistic trajectories. If the atomic COM is treated as quantum, or when there is no common rest frame, this additional term cannot be neglected. We will also discuss the higher order terms that appear in the case of more relativistic trajectories of the COM.

We compare these considerations to the usually employed effective light-matter interaction models.
Thus, we discuss the limitations of the effective dipolar coupling $e\hat{\bm r}\cdot\hat{\bm E}(\hat{\bm r})$ and scalar-analogue models such as the Unruh-DeWitt model.
In the case of scalar-analogue models, we argue here that a coupling of COM and radiation degrees of freedom has to be included in most scenarios if one wants to capture the atomic dynamics. 

Finally, we will show that considering only the effective dipole term for a classical COM still yields relativistically covariant predictions. We will provide arguments that, even if we are failing to describe precise atomic physics with this simplification, there is utility in using this interaction as a testbed to implement measurements on the electromagnetic field whose qualitative behavior captures the features of the light-matter interaction under some assumptions.

The manuscript is organized as follows: in Sec.~\ref{sec effective models} we present two of the common effective light-matter models, namely the Unruh-DeWitt and the effective dipolar coupling model. We discuss the assumptions of these models and their consequent limitations.
In Sec.~\ref{sec multipolar} we will re-derive the multipolar Hamiltonian at the level of the Schr\"odinger equation.  In this section we will always work with the quantum electromagnetic field, giving a position-representation of the multipolar Hamiltonian in its dipole approximation in terms of the internal hydrogenic wavefunctions and external COM eigenstates. Further, we are going to show the impact of COM dynamics and R\"ontgen term at the example of transition rates. 
We will then discuss in Sec.~\ref{sec rel cor} the impact of leading order relativistic corrections.
In Sec.~\ref{sec eff dipole 2}, we revisit the effective dipole model to show its qualitative merits in relativistic scenarios and its covariance under Lorentz transformations.
In Sec.~\ref{sec modify udw} we are proposing modifications for the Unruh-DeWitt model under the considerations of the previous sections to account for COM dynamics.

\section{Effective light-matter models}\label{sec effective models}

\subsection{The Unruh-DeWitt model}\label{sec effective UDW}

In the context of RQI, or generally if the objective is to obtain information in a QFT setting, the notion of a \textit{particle detector} that can extract these information locally from a quantum field is crucial. A particle detector is an internally non-relativistic quantum system that couples in a covariant way to a second-quantized field. It circumvents the problems of projective measurements in QFT \cite{sorkin, dowker}, and may give rise to a phenomenological interpretation for the elusive notion of particles in QFT \cite{kuhlmann}. Particle detectors have been crucially used in a plethora of scenarios in quantum field theory in flat and curved spacetimes (e.g., the Unruh and Hawking effects \cite{PhysRevD.14.870, hawkingeff}, cosmological particle creation \cite{gibbons}, entanglement harvesting \cite{valentini, pozas2}, etc). The most common model of a particle detector is the Unruh-DeWitt (UDW) model, e.g. \cite{PhysRevD.14.870, deWitt}. This model typically considers a two-level non-relativistic quantum system rigidly localized in space and time that covariantly couples to a quantum scalar field amplitude $\hat\phi(t,\bm x)$ along its (possibly relativistic) trajectory. The UDW interaction-picture interaction Hamiltonian in the most general case is given by~\cite{tales1}:
\begin{align}
    \hat H_\textsc{udw} =\lambda \chi(\tau)  \hat{\mu}(\tau)\otimes\int_{\Sigma_{\tau}}\!\! \mathrm{d}^{3} \bm\xi \sqrt{-g}\,  F(\bm \xi) \hat{\phi}(t(\tau,\bm{\xi}),\bm x(\tau,\bm{\xi})),
\end{align}
where $[t,\bm x]$ is the field quantization frame,  $[\tau,\bm\xi]$ is the Fermi-Walker frame comoving with the center-of-mass of the detector, $\Sigma_\tau$ are the spatial sections associated with the coordinates $[\tau,\bm\xi]$, $g$ is the determinant of the metric, $\hat\mu(\tau)$ is the monopole moment representing the internal degree of freedom of the detector, $\chi(\tau)$ encodes the time-dependence of the coupling in the detectors COM frame, $F(\bm{\xi})$ is the spatial profile of the detector, and, finally, $\lambda$ is the coupling strength.

In flat spacetime, and for a detector comoving with the field quantization frame, this Hamiltonian simply becomes, by identifying $\tau=t$ and $\bm\xi=\bm x$,
\begin{align}
\hat H_\textsc{udw} 
= \lambda\, \chi(t)\hat{\mu}(t)\otimes \int_{\mathbb{R}^3}\dd^{3}\bm{x} \  F(\bm{x}) \, \hat{\phi}(t,\bm{x}).\label{UDW}
\end{align}
Although simple, this Hamiltonian already captures a large amount of the phenomenology of the light-matter interaction. Indeed, the popular Dicke \cite{dickemodel} and Jaynes-Cummings model~\cite{Scully} are but further simplifications of the Unruh-DeWitt model (typically assuming pointlike detectors, single mode approximation and some form of rotating wave approximation).

The power of the Unruh-DeWitt model lies in its computational applicability: while it certainly gives a reasonable effective model  to carry out measurements on quantum fields, computable results can be obtained  even in complicated curved spacetime scenarios or involved relativistic detector trajectories. 

There are, however, shortcomings of the model when it comes to describing the light-matter interaction. First, the scalar nature of the coupling makes it impossible for the model to capture phenomenology associated with the exchange of angular momentum between the detector and the field. Also, the spatial smearing has to be prescribed `by hand' since we do not have a first-principle-inspired reason to choose the exact shape of the detector's localization. Finally, this model considers that the center of mass of the detector is a classical degree of freedom whose dynamics is decoupled from the detector's internal levels. This does not mean that the model is not useful, but rather that refinements are needed if we want to go beyond rough order of magnitude estimates in realistic atomic systems, or in regimes where the neglected aspects of the interaction play a key role.


\subsection{The dipole coupling}\label{sec effective dipole}

One step forward in adding complications to the effective light-matter interaction models is obtained by assuming that the atom is modelled by a classical infinite mass proton (as compared to the electron) that generates a classical Couloumb potential in the atomic COM frame (which generates the internal energy levels for the atomic system). Then, the atom couples dipolarly to a time dependent second-quantized electric field as seen from the COM frame of the detector:
    \begin{align}
    \hat H_{\text{eff}}&= \hat H_0 +\hat H_I,\\
    \hat H_0&=\frac{\hat{\bm{p}}_e^2}{2 m_e}-\frac{1}{4 \pi\epsilon_0 } \frac{e^{2}}{|\hat{\bm{r}}_e|},\\
    \hat H_I&=e \hat{\bm r}_e\cdot \hat{\bm E}(t,\hat{\bm r}_e),\label{eff dip}
    \end{align}
 where, for simplicity, we assumed that the atom is comoving with the field quantization frame (something that we will relax in Sec.~\ref{sec eff dipole 2}). To effectively compare this model with the Unruh-DeWitt model, let us introduce a position representation in terms of the hydrogenic wavefunctions that are solutions of the Schr\"odinger equation for $\hat H_0$. That is, $\Psi_{\bm{a}}(\bm r_e)=\braket{\bm r_e}{\bm{a}}$, where $\{\ket {\bm{a}}=\ket{(n,l,m)}\}$  such that~\cite{Martin-MartinezMOnteroDelRey, pozas}
\begin{align}
  \hat H_I=& e\sum_{\bm{a}, \bm{b}} \int_{\mathbb{R}^3} \dd^3 \bm  r_e\braket{\bm{a}}{\bm r_e}\braket{\bm r_e}{\bm{b}}\bm{r_e}\cdot \hat E(t,\bm r_e)  \dyad{\bm{a}}{\bm{b}} \nonumber\\
   =&e\sum_{\bm{a},\bm{b}} \int_{\mathbb{R}^3} \dd^3 \bm  r_e\,\Psi_{\bm{a}}(\bm r_e)^*\Psi_{\bm{b}}(\bm r_e)\bm{r_e} \cdot \hat E(t,\bm r_e) \dyad{\bm{a}}{\bm{b}}\nonumber\\
   \eqqcolon& \sum_{\bm{a}> \bm{b}} \int_{\mathbb{R}^3}\dd^3 \bm  r_e\, \hat{\bm{d}}_{\bm{a}\bm{b}}(\bm r_e)\cdot \hat E(t,\bm r_e), \label{eff pos}
\end{align}
where in the last step we defined the dipole operator $\hat{\bm{d}}_{\bm{a}\bm{b}}(\bm r_e)$,  and the ordering $\bm{a}>\bm{b}$ is first with respect to $n$, then $l$ and lastly $m$---so as to follow the energy hierarchy approximately, (although in this approximate model only the quantum number $n$ gives the internal energy of the atom). Notice that the diagonal terms of the dipole operator can be directly removed since there is a change of parity selection rule for electric dipole transitions~\cite{feuerbacher}.

If we express the internal atomic degrees of freedom in the interaction picture with respect to time $t$ as well, the dipole operator between two levels $\ket{\bm a}$ and $\ket{\bm b}$ is of the form
\begin{align}
     \hat{\bm{d}}_{\bm{a}\bm{b}}(t,\bm r_e)=e \bm F_{\bm{a}\bm{b}}(\bm r_e) e^{\ii \Omega_{\bm{a}\bm{b}} t}\dyad{\bm{a}}{\bm{b}} + \text{H.c.}\label{dip form}
\end{align}
The spatial \textit{smearing vector} is given by the hydrogen wavefunctions of the two levels connected by each matrix element:  \mbox{$\bm F_{\bm{a}\bm{b}}(\bm r_e)=\bm r_e\Psi_{\bm{a}}^*(\bm r_e)\Psi_{\bm{b}}(\bm r_e)$}, and $\hbar\Omega_{\bm{a}\bm{b}}$ is the energy difference between the states $\ket{\bm{a}}$ and $\ket{\bm{b}}$.
    

In contrast to the UDW model where the spatial localization of the coupling was introduced by hand, from \eqref{dip form} we see that the localization of the dipolar interaction is governed by the electronic wavefunctions. 
In this light, when we add a switching function modelling the beginning and the end of a finite-time process, a comparison of equations \eqref{eff pos} and \eqref{UDW} shows in what sense this model is a refinement of the UDW model for the light-matter interaction: we could think of the Unruh-DeWitt coupling as the scalar version of this effective dipole coupling, and we have a way to prescribe the localization of the coupling out of the physical assumptions of the dipolar model without having to introduce it \textit{ad-hoc}.

The advantages of the effective dipole coupling are that it is still a simple model, as the only quantum degree of freedom of the atom is the position of the electron. Furthermore, it still allows for arbitrary relativistic trajectories for the COM frame, whose position is treated classically (as we will see in Sec.~\ref{sec eff dipole 2}). Additionally it does allow for the exchange of orbital angular momentum between the detector internal degrees of freedom and the electromagnetic field. Also, the dipole coupling is a) gauge unambiguous, and b) it is inspired by typical light-matter interaction assumptions where higher multipoles are neglected. 

 However, this is still an effective model. We emphasize again that the assumptions that went into the derivation of \eqref{eff dip} neglect the dynamics of any atomic degrees of freedom other than the ones associated with the electron. In that sense, the dipole term is introduced somewhat \textit{ad hoc}, instead of rigorously obtained from the two-particle minimal-coupling light-matter interaction after careful gauge and multipole considerations are taken into account. Same as the UDW model, this does not mean that the model is not useful. In fact, as we will discuss in Sec.~\ref{sec eff dipole 2} this model can also be made fully covariant same as it was shown for the Unruh-Dewitt model in \cite{tales1,Pablo1}. Rather, we argue that one has to refine this model if one wants to go beyond qualitative results and rough order of magnitude estimations, and instead one wants to predict outcomes of experiments in more involved regimes where the assumptions of the model are not fulfilled.

\section{The multipolar coupling Hamiltonian}\label{sec multipolar}

Our objective in this section is to explicitly derive the multipolar coupling Hamiltonian from the two-particle minimal coupling. We will do so for a fully quantized model---including the quantization of both the field and center of mass of the atom.  More concretely, we combine the quantization of the field, the COM of the atom and the relative motion (internal) degree of freedom to derive the dipole coupling Hamiltonian in the (approximated) gauge in which the relative degree of freedom wavefunctions correspond exactly to the textbook-problem of a charge trapped in a Coulomb potential (hydrogenoid atom). It is important to recall that  that the atomic wavefunctions are not gauge-invariant~\cite{Scully,funai}, and only under  very strict considerations the internal atomic wavefunctions are the textbook hydrogen-like ones. 

Although we are (to a large extent) revisiting old-known problems, the particular approach to deriving these results from the Hamiltonian formalism with a fully quantum framework that we take is (to the authors' knowledge) not available in previous literature. Operating directly from the Hamiltonian formalism we avoid introducing an \textit{ad-hoc} change of the canonical commutation relations of the field operators, something common in past derivations of the multipolar Hamiltonian  (e.g., \cite{Babiker83,Babiker84,Babiker93}), which allows for a pedagogically easier treatment. We will also analyze all the terms that are typically neglected in simplified particle detector models employed in RQI, such as the orbital magnetic dipole and R\"ontgen terms~\cite{Babiker84,Babiker93,Wilkens93}, paying special attention to the discussion about gauge and localization. 

We consider a hydrogen atom interacting with the electromagnetic field. We will treat the electromagnetic field as a second quantized system and the internal structure of the atom as a first quantized system. The electromagnetic field is described by the gauge-dependent potential operators $(\hat U, \hat{\bm A})$. The atom consists of a proton with mass $m_p$ and associated position operator $\hat{\bm r}_p$ and similarly an electron with mass $m_e$ and position operator $\hat{\bm r}_e$. Both constituents will be treated as spinless.

 A relativistic-friendly approach would start from the general classical Lagrangian with the minimal coupling prescription 
 \begin{align}
     L=&-\sum_{i=e,p} m_{i} c^{2} \sqrt{1-\dot{\bm{r}}_{i}^{2} / c^{2}}\nonumber\\
     &+\frac{\epsilon_{0}}{2} \int_{\mathbb{R}^3} \dd^{3} \bm{x}\left(\left(\partial_{t} \bm{A}_{\text{tot}}+\nabla U\right)^{2}-c^{2}\left(\nabla \times \bm{A}_{\text{tot}}\right)^{2}\right)\nonumber\\
     &+\int_{\mathbb{R}^3} \dd^{3} \bm{x}\left(\bm{j} \cdot \bm{A}_{\text{tot}}-\rho U\right),
 \end{align}
 where $\bm A_{\text{tot}}$ includes the vector potential generated by the charges, and $\rho$, $\bm j$ are the charge and current densities, respectively.
 Solving the dynamics for this Lagrangian is involved so that generally one is reduced to an expansion about the particle velocities $\dot{\bm{r}}_i$ in some inertial frame. Changing to the Hamiltonian picture and after quantization we get the minimal coupling Hamiltonian at leading order in velocities. Besides the standard free-field Hamiltonian, this reads~\cite{Schleich}
\begin{align}
  \hat H=&\sum_{i={e,p}}\left[\frac{(\hat{\bm p}_i + e_i \hat{\bm A}(t,\hat{\bm r}_i))^2}{2m_i} -e_i \hat U(t,\hat{\bm r}_i)\right]-\frac{e^2}{4\pi\epsilon_0|\hat{\bm r}|}\label{H start},
\end{align}
where the last term corresponds to the electrostatic Coulomb energy with $\hat{\bm r}=\hat{\bm r}_e-\hat{\bm r}_p$, and we are considering the field in the interaction picture with explicit time dependence.
 The sub-leading relativistic correction, called the Darwin Hamiltonian~\cite{barnett2018}, is of the form  
 \begin{align}
    \hat H_{\text{D}}=&\frac{\hat{\bm{\pi}}_{e}^{4}}{8 m_{e}^{3} c^{2}}+\frac{\hat{\bm{\pi}}_{p}^{4}}{8 m_{p}^{3} c^{2}}+\frac{e^2}{16 \pi \varepsilon_{0} c^{2} m_{e} m_{p}}\nonumber\\
    &\times\left[\hat{\bm{\pi}}_{e} \cdot \frac{1}{|\hat{\bm r}|} \hat{\bm{\pi}}_{p}+\left(\hat{\bm{\pi}}_{e} \cdot \hat{\bm{r}}\right) \frac{1}{|\hat{\bm r}|^{3}}\left(\hat{\bm{r}} \cdot \hat{\bm{\pi}}_{p}\right)+(e \leftrightarrow p)\right],\label{darwin}
    \end{align}
    where $\hat{\bm{\pi}}_{i}:=\hat{\bm{p}}_{i}+e_{i} \hat{\bm{A}}\left(\hat{\bm{r}}_{i}\right)$.
 In Sec.~\ref{sec modify udw}, once we derived the dipolar Hamiltonian, we will come back and discuss the phenomenoligcal implications of $ \hat H_{\text{Darwin}}$.
 
For simplicity, in this section we shall be concerned with general scenarios where the atomic COM describes non-relativistic motion. This means that here we will neglect the Darwin term and any other higher order corrections associated with the relativistic motion of the charges.  While this is not covering all interesting regimes in RQI, it does cover  several relevant regimes directly such as, for instance, most entanglement harvesting scenarios~\cite{valentini,pozas2,smith}. We will leave the discussion of regimes with relativistic atomic motion for Sec.~\ref{sec rel cor}.  

We can therefore start from the standard leading-order minimal coupling Hamiltonian in Eq.~\eqref{H start}. For convenience, we choose the Coulomb gauge where there is no scalar potential and $[\hat{\bm p}_i,\hat{\bm A}(t,\hat{\bm r}_i)]=0$ \cite{Scully,cohen}.

When working with the minimal coupling Hamiltonian we have to be careful with the gauge freedom of the field. In particular, we need to make a consistent choice of atomic wavefunctions when we choose a particular gauge in order to have gauge-independent predictions. For example, in the Coulomb gauge, the atomic wavefunctions of a hydrogen atom are very different from the textbook hydrogen orbitals (see e.g., \cite{lamb,Scully,funai}).

Additional complications appear as we are working with a two-particle system. We cannot simply assume that there is a gauge where the internal atomic wavefunctions are the textbook atomic orbitals and then transform them to whatever gauge we are considering. As we will see, there is no such gauge. 
Moreover, in general one cannot directly neglect the $\hat{\bm A}^2$ terms. This is only possible in a few certain regimes most of them outside of the scope of RQI setups (see, e.g.,~\cite{asquare}).

It would be convenient to express the Hamiltonian solely in terms of gauge-invariant field observables, and also choose canonical coordinates so that we have the hydrogenic orbitals when we take the position representation for the relative motion degree of freedom for the atom.
The canonical transformation that achieves these two goals is a Power-Woolley-Zienau (PZW) transformation~\cite{Babiker93}. This transformation applied to the Coulomb-gauge Hamiltonian yields the so-called multipolar coupling Hamiltonian. 

Concretely, let us define the atomic center-of-mass and relative motion position operators:
\begin{align}
    \hat{\bm R}=\frac{m_e \hat{\bm r}_e +m_p \hat{\bm r}_p}{M}, \quad \hat{\bm r}=\hat{\bm r}_e-\hat{\bm r}_p,
\end{align}
where $M=m_e+m_p$.
Similarly the total momentum of the center-of-mass, and the momentum of the relative motion associated with the reduced mass $\mu=m_e m_p/M$ read, respectively
\begin{align}
  \hat{\bm P}=\hat{\bm p}_e+\hat{\bm p}_p,\quad \hat{\bm p}=\frac{m_p}{M}\hat{\bm p_e}-\frac{m_e}{M}\hat{\bm p_p}.  
\end{align}
These two new sets of operators satisfy the canonical commutation relations: $[\hat{\bm R},\hat{\bm P}]=\ii \hbar\openone=[\hat{\bm r},\hat{\bm p}]$.
The Hamiltonian \eqref{H start} re-expressed in terms of center-of-mass and relative coordinates yields
\begin{align}
    \hat H&=\frac{\hat{\bm{P}}^{2}}{2 M}+\frac{\hat{\bm{p}}^{2}}{2 \mu}-\frac{1}{4 \pi\epsilon_0 } \frac{e^{2}}{|\hat{\bm{r}}|}\nonumber\\*
    &-\frac{e}{\mu}\!\left\{\!\frac{\mu}{m_{e}} \hat{\bm{A}}\left(t,\hat{\bm{R}}+\frac{m_{p}}{M} \hat{\bm{r}} \right)+\frac{\mu}{m_{p}} \hat{\bm{A}}\left(t,\hat{\bm{R}}-\frac{m_{e}}{M} \hat{\bm{r}}\right)\!\right\} \!\cdot\hat{\bm{p}}\nonumber\\*
    &-\frac{e}{M}\left\{\hat{\bm{A}}\left(t,\hat{\bm{R}}+\frac{m_{p}}{M} \hat{\bm{r}}\right)-\hat{\bm{A}}\left(t,\hat{\bm{R}}-\frac{m_{e}}{M} \hat{\bm{r}}\right)\right\} \cdot\hat{\bm{P}}\nonumber\\*
    &+\frac{e^{2}}{2 m_{e}} \hat{\bm{A}}^{2}\left(t,\hat{\bm{R}}+\frac{m_{p}}{M} \hat{\bm{r}}\right)+\frac{e^{2}}{2 m_{p}} \hat{\bm{A}}^{2}\left(t,\hat{\bm{R}}-\frac{m_{e}}{M} \hat{\bm{r}}\right).\label{H neu}
\end{align}
The non-relativistic quantum treatment of the atom requires the center-of-mass and relative momenta to be bounded. Since the motion of an electron `around' a proton is typically non-relativistic, considering   for the relative motion to be non-relativistic is generally a very reasonable assumption.  However, for the state of the COM of the atom to be in a non-relativistic regime, the state should not have any non-negligible overlap with generalized eigenstates of momentum beyond some scale, where relativistic corrections would be necessary.

In order to arrive at the multipolar Hamiltonian, we insert resolutions of identity in the COM and relative position bases (taking a position representation for $\bm R$ and $\bm r$), and expand the vector field around the center-of-mass coordinate $\bm R$. For our purposes, we will only consider the dipolar contributions:
\begin{align}
    \hat{\bm A}(t,\bm R+\delta \bm r) \approx  \hat{\bm A}(t,\bm R)+ (\delta\bm r \cdot \bm\nabla_{\bm R})\hat{\bm A}(t,\bm R).\label{dipole expansion}
\end{align}
When applied to Eq.~\eqref{H neu}  we wil have that either \mbox{$\delta\bm r=\frac{m_{p}}{M} \bm{r}$} or  $\delta\bm r=-\frac{m_{e}}{M} \bm{r}$ depending on the term. As we will discuss more in depth later on, the spatial support in the relative coordinate $\bm r$ for atomic scales is given approximately by the scale of Bohr radius $a_0$ associated with the reduced mass $\mu$.
Hence, the second-order term is suppressed with respect to the leading order by a factor $\sim a_0 |\bm k_{\textsc{uv}}|$, with  $|\bm k_{\textsc{uv}}|$ being the maximum wave vector of the vector field. It may be determined by the atomic smearing and the time-dependent coupling between atom and field, or by a dominant atomic transition process \cite{funai}. Ultimately, the Compton wavelength will yield the upper bound in order to stay in the non-relativistic quantum description of the atom. Note, that since we consider a quantum COM, or also in the case of motion of a classical COM, the second order term is required even at the dipole level.

In the dipole regime where  Eq.~\eqref{dipole expansion} applies, $\hat H$ approximates to 
\begin{align}
    \hat{H}^{(1)}=&\int_{\mathbb{R}^3} \dd^3 \bm  R \int_{\mathbb{R}^3} \dd^3 \bm  r \left\{\frac{1}{2 M}\left[\hat{\bm{P}}-e\left(\bm{r} \cdot \bm{\nabla}_{\bm{R}}\right) \hat{\bm{A}}(t,\bm{R})\right]^{2}\right.\nonumber\\
    &+\frac{1}{2 \mu}\left[\hat{\bm{p}}-e \hat{\bm{A}}(t,\bm{R})-e \frac{\Delta m}{M}\left(\bm{r} \cdot \bm{\nabla}_{\bm{R}}\right) \hat{\bm{A}}(t,\bm{R})\right]^{2} \nonumber\\
    & \left.-\frac{e^2}{4\pi\epsilon_0|\bm r|}\right\}\dyad{\bm R}{\bm R}\otimes\dyad{\bm r}{\bm r},\label{H(1)}
\end{align}
where $\Delta m=m_p-m_e$.
The Hamiltonian \eqref{H(1)} is the generator of time translations to the joint atom-field state $\ket\Psi$ governed by the Schr\"odinger equation
\begin{align}
    \ii\hbar \pdv{\ket{\Psi}}{t}=\hat H^{(1)} \ket{\Psi},\label{sg}
\end{align}
with the field being in the interaction picture.
We will now write the interaction Hamiltonian in terms of gauge-invariant field operators, and such that the internal atomic Hamiltonian admits the usual hydrogen wavefunction solutions. To accomplish this, we perform a local canonical transformation generated by the self-adjoint operator~\cite{Schleich}
\begin{align}
     \hat\Lambda^{(1)}(t,\hat{\bm{R}},\hat{\bm{r}})&=\int_{\mathbb{R}^3}\dd^3 \bm  R\,\dyad{\bm R}{\bm R}\left[\hat{\bm{r}}\cdot \hat{\bm{A}}(t,\bm{R})\right.\nonumber\\
     &\quad\left.+\frac{\Delta m}{2M}(\hat{\bm r} \cdot \bm\nabla_{\bm R})\left(\hat{\bm r}\cdot\hat{\bm A}(t,\bm R)\right)\right].\label{gauge 1}
\end{align}
This transformation is, in general, not a gauge transformation. We will see later in Sec.~\ref{sec eff dipole 2} that for the effective dipole model one can indeed use a gauge transformation to go from one-particle minimal coupling to the multipolar Hamiltonian, but not in the current two-particle case.
Note that the procedure of first performing the dipole approximation \eqref{dipole expansion} and then performing the canonical transformation $\hat U_{\hat\Lambda^{(1)}}\coloneqq \text{exp}[-\frac{i}{\hbar} e \hat\Lambda^{(1)}]$ is equivalent to first performing a  transformation with the Dirac-Heisenberg line function
\begin{align}
    \hat\Lambda(t,\hat{\bm R},\hat{\bm r}) = \hat{\bm r} \cdot\int_{0}^{1} \dd \lambda \, \hat{\bm{A}}\left(t,\hat{\bm R}-\left(\frac{m_e}{M}-\lambda\right) \hat{\bm r}\right),\label{gauge 2}
\end{align}
and then performing a Taylor expansion in the electromagnetic vector potential~\cite{Schleich}. Furthermore \eqref{gauge 2} is identical (order by order) to the standard PZW transformation \cite{Babiker93} (as we show in Appendix~\ref{pzw check}):
\begin{align}
     \hat\Lambda^{\text{PZW}}=& \sum_{i=e,p} \frac{e_i}{|e|} (\hat{\bm r}^i-\hat{\bm R}) \cdot \int_{0}^{1} \dd \lambda\,  \hat{\bm{A}}\left(t,\hat{\bm R}+\lambda\big(\hat{\bm r}^i- \hat{\bm R}\big)\right)\label{pzw gauge}.
\end{align}
Using \eqref{gauge 1} we define the canonically transformed state $\tilde{\ket{\Psi}}$ through
\begin{align}
    \ket{\Psi}=\exp(\frac{\ii}{\hbar} e  \hat\Lambda^{(1)}(t,\hat{\bm{R}},\hat{\bm{r}}))\tilde{\ket{\Psi}}.\label{trafo}
\end{align} 
This means that the left-hand side of \eqref{sg} can be written as
\begin{align}
   \ii \hbar \pdv{\ket{\Psi}}{t}&=-e \pdv{\hat\Lambda^{(1)}}{t}e^{\frac{\ii}{\hbar} e \hat\Lambda^{(1)}}  \tilde{\ket{\Psi}} + e^{\frac{\ii}{\hbar} e \hat\Lambda^{(1)}}\ii\hbar \pdv{\tilde{\ket{\Psi}}}{t},
\end{align}
while the right-hand side of \eqref{sg} can be written as
\begin{align}
\hat H^{(1)} \ket{\Psi}=\hat H^{(1)}  \exp\left(\frac{\ii}{\hbar} e \hat\Lambda^{(1)}\right) \tilde{\ket{\Psi}}.
\end{align}
Regrouping all the extra terms in the left-hand-side into the right-hand side allows us to see the form of the canonically transformed Hamiltonian 
\begin{align}
 \hat{\tilde{H}}^{(1)}=\exp\left(-\frac{\ii}{\hbar} e \hat\Lambda^{(1)}\right)   \!\left[\hat H^{(1)}+e \pdv{\hat\Lambda^{(1)}}{t}\right] \! \exp\left(\frac{\ii}{\hbar} e \hat\Lambda^{(1)}\right).\label{new H}
\end{align}
As we will see later, $\hat{\tilde{H}}^{(1)}$ will be the Hamiltonian we are seeking: a function of the electric and magnetic field operators, and for which the internal atomic dynamics admits as solution the textbook hydrogen wavefunctions. Notice that the canonically transformed (PZW-transformed)  Hamiltonian is not unitarily equivalent to the minimal coupling Hamiltonian (after the dipole approximation). As we will discuss later the extra term (associated with the time-dependence of $\hat\Lambda$) is related with self-energy and will be responsible for a shift on the energy levels (such as the Lamb shift).

To implement this canonical transformation, we need the commutation relations between the vector potential and its different derivatives. In terms of the usual plane-wave expansion, the vector potential in the interaction picture takes the form
\begin{align}
	\hat{\bm A}(t,\bm x)&=\sum_{s=1}^2 \int_{\mathbb{R}^3} \frac{\text{d}^3k}{(2 \pi)^{\frac{3}{2}}} \sqrt{\frac{\hbar }{2 \epsilon_0 c |\bm k|}} \left(  \hat{a}_{\bm k, s} \bm \epsilon_{\bm k, s} e^{\ii  \mathsf{k} \cdot  \mathsf{x}} + \text{H.c.} \right),
	\end{align}
	where $\mathsf{k}$ and $\mathsf{x}$ are respectively the  four-wavevector and four-position four-vectors, and we work with the metric signature $(-,+,+,+)$. We denoted as $\{\bm \epsilon_{\bm k, s}\}, s\in{1,2}$  an arbitrary set of two orthonormal transverse polarization vectors that together with the normalized wave vector $\bm e_{\bm k}=\bm k/|\bm k|$ form an orthonormal basis in $\mathbb{R}^3$. Therefore we find that the equal-time commutator between two  components of the vector potential is
\begin{align}
	    \left[ 	\hat{A}^i(t,\bm x),	\hat{A}^j(t,\bm x')\right]=&\int_{\mathbb{R}^3} \!\frac{\text{d}^3k}{(2 \pi)^{3}} \frac{\hbar }{2 \epsilon_0 c |\bm k|}    (\delta^{ij} -  e^i_{\bm k}e^j_{\bm k}) \nonumber\\
	    &~~\cdot\left( e^{\ii  \bm{k} \cdot (\bm{x}-\bm{x}')}-e^{-\ii  \bm{k} \cdot (\bm{x}-\bm{x}')} \right), \label{comm 1}
	\end{align}
	by use of the completeness relations~\cite{cohen}
	\begin{align}
   \sum_{s=1}^2  \epsilon_{\bm k, s}^i   \epsilon_{\bm k, s}^j = \delta^{ij} -  e^i_{\bm k}e^j_{\bm k}. \label{completeness}
\end{align}	
	By differentiation, we find the remaining commutators required for the dipole approximation (we can stop at the first spatial derivatives). The details of the calculations can be found in Appendix~\ref{commutators}.
	We use the transverse delta function~\cite{delta}
	\begin{align}
	\delta^{i j, (\text{tr})}(\bm x)=\frac{1}{(2\pi)^3}\int_{\mathbb{R}^3} \dd^3 \bm  k\, (\delta^{ij} -  e^j_{\bm k}e^j_{\bm k}) e^{\ii \bm k\cdot \bm x}.
	\end{align}
		The only commutators that are non-zero in the coincidence limit are then
	\begin{align}
	       \left[ 	\hat{A}^i(t,\bm x),	\partial_{t} \hat{A}^j(t,\bm x')\right]=&\frac{\ii \hbar}{\epsilon_0} \delta^{i j, (\text{tr})}(\bm x-\bm x'),  \label{comm 4} \\
	    \left[ \partial_l	\hat{A}^i(t,\bm x),	\partial_{t}\partial_m \hat{A}^j(t,\bm x')\right]=&\frac{\ii \hbar}{\epsilon_0} \pdv{\delta^{i j, (\text{tr})}(\bm x-\bm x')}{x^l}{x'^m}.  \label{comm 7}
	\end{align}
	Eq.~\eqref{comm 4} and \eqref{comm 7} contribute to the commutator of the generator $\hat\Lambda^{(1)}$ with its time derivative. Moreover, they yield divergent contributions in the coincidence limit which will give rise to the self-energy of the atom. They appear only in the quantum case and its divergences can be renormalized and regularized through smeared spatial profiles.

To find the new Hamiltonian \eqref{new H}, we commute the old Hamiltonian with the canonical transformation operator. There will be two kinds of contributions: those that come from commuting with $\hat H^{(1)}$ and those that come from commuting with $\partial_t\hat\Lambda$. Since the calculation can get \mbox{cumbersome}, let us compute the two non-trivial terms in $\hat H^{(1)}$  as well as the contributions from the commutator with $\partial_t\hat\Lambda$ separately.

First the commutation of the canonical transformation with the first summand  of $\hat H^{(1)}$ in Eq.~\eqref{H(1)}. To that end, let us consider initially the simpler commutation (without the square) given by  
\begin{align}
    &\int_{\mathbb{R}^3} \dd^3 \bm  R\dyad{\bm R}{\bm R}\left[\hat{\bm{P}}-e\left(\hat{\bm{r}} \cdot \bm{\nabla}_{\bm{R}}\right) \hat{\bm{A}}(t,\bm{R})\right]\hat U_{\hat\Lambda^{(1)}}^\dagger \tilde{\ket{\Psi}}\nonumber\\
    =&\int_{\mathbb{R}^3} \dd^3 \bm  R \dyad{\bm R}{\bm R}U_{\hat\Lambda^{(1)}}^\dagger\nonumber\\
    &\quad\times\left[\hat{\bm{P}}+e\bm\nabla_{\bm R}\hat\Lambda^{(1)} -e\left(\hat{\bm{r}} \cdot \bm{\nabla}_{\bm{R}}\right) \hat{\bm{A}}(t,\bm{R})\right] \tilde{\ket{\Psi}},
\end{align}
with $\bm\nabla_{\bm R}\hat{\Lambda}^{(1)}(t,\bm R,\bm r)= \bm{\nabla}_{\bm{R}}\big[\bm{r} \cdot \hat{\bm{A}}(t,\bm{R})]$ to leading order~\cite{Schleich}. This term arises from position representation, i.e. $\bra{\bm R}\hat{\bm P}\hat O\ket{\Psi}=-\ii\hbar \bm\nabla_{\bm R}\bra{\bm R}\hat O\ket{\Psi}$.
Using
\begin{align}
    \hat{\bm B}&=\bm\nabla \times \hat{\bm A}\label{B from A},\\
    \hat{\bm r} \times \hat{\bm B}&=\bm\nabla(\hat{\bm r} \cdot \hat{\bm A})-(\hat{\bm r}\cdot \bm\nabla )\hat{\bm A}\label{help 2},
\end{align}
and recovering the square, we arrive at
\begin{align}
    &\int_{\mathbb{R}^3} \dd^3 \bm  R\dyad{\bm R}{\bm R}\left[\hat{\bm{P}}-e\left(\hat{\bm{r}} \cdot \bm{\nabla}_{\bm{R}}\right) \hat{\bm{A}}(t,\bm{R})\right]^2 U_{\hat\Lambda^{(1)}}^\dagger\tilde{\ket\Psi}\nonumber\\
    &=U_{\hat\Lambda^{(1)}}^\dagger \left[\hat{\bm P}+ e \hat{\bm r}\times \hat{\bm B}(t,\hat{\bm R})\right]^2\tilde{\ket{\Psi}}.\label{cm momentum}
\end{align}
Similarly, for the next summand of $\hat H^{(1)}$, we need
\begin{align}
    &\int_{\mathbb{R}^3} \dd^3 \bm  R  \int_{\mathbb{R}^3} \dd^3 \bm  r\dyad{\bm R}{\bm R}\otimes \dyad{\bm r}{\bm r}\nonumber\\*
    & \times\left[\hat{\bm{p}}-e \hat{\bm{A}}(t,\bm{R})-e \frac{\Delta m}{M}\left(\bm{r} \cdot \bm{\nabla}_{\bm{R}}\right) \hat{\bm{A}}(t,\bm{R})\right]U_{\hat\Lambda^{(1)}}^\dagger\tilde{\ket{\Psi}}\nonumber\\*
    &=\int_{\mathbb{R}^3} \dd^3 \bm  R \int_{\mathbb{R}^3} \dd^3 \bm  r \dyad{\bm R}{\bm R}\otimes\dyad{\bm r}{\bm r}U_{\hat\Lambda^{(1)}}^\dagger\left[\hat{\bm{p}}-e \hat{\bm{A}}(t,\bm{R})\right.\nonumber\\*
    &\quad\left.-e \frac{\Delta m}{M}\left(\bm{r} \cdot \bm{\nabla}_{\bm{R}}\right) \hat{\bm{A}}(t,\bm{R})+ e\bm\nabla_{\bm r}\hat{\Lambda}^{(1)}\right]\!\tilde{\ket{\Psi}},
\end{align}
where, using Eq.~\eqref{gauge 1}, we get
\begin{align}
  \nonumber\  \bm\nabla_{\bm r}\hat{\Lambda}^{(1)}(t,\bm R,\bm r)=&\hat{\bm{A}}(t,\bm{R})+ \frac{\Delta m}{2M}\left\{ \bm\nabla_{\bm R}\left[\bm r\cdot\hat{\bm A}(t,\bm R)\right]\right.\\
    &\left.+ (\bm r\cdot\bm\nabla_{\bm R})\hat{\bm A}(t,\bm R)\right\}.
\end{align}
Thus, recovering the square, and using \eqref{B from A} and \eqref{help 2}, we have
\begin{align}
    &\int_{\mathbb{R}^3} \dd^3 \bm  R \int_{\mathbb{R}^3} \dd^3 \bm  r\dyad{\bm R}{\bm R}\otimes\dyad{\bm r}{\bm r} \nonumber\\
    &\quad\times\left[\hat{\bm{p}}-e \hat{\bm{A}}(t,\bm{R})-e \frac{\Delta m}{M}\left(\bm{r} \cdot \bm{\nabla}_{\bm{R}}\right) \hat{\bm{A}}(t,\bm{R})\right]^2 U_{\hat\Lambda^{(1)}}^\dagger\tilde{\ket\Psi}\nonumber\\
    &=U_{\hat\Lambda^{(1)}}^\dagger\left[\hat{\bm{p}}+\frac{e}{2} \frac{\Delta m}{M}(\hat{\bm{r}} \times \hat{\bm{B}}(t,\hat{\bm{R}}))\right]^{2}\tilde{\ket{\Psi}}.\label{rel momentum}
\end{align}
This concludes the calculations regarding $\hat H^{(1)}$ as the Coulomb potential stays trivially the same.
In the last step to find the new Hamiltonian, we have to evaluate  $U_{\hat\Lambda^{(1)}}(\partial_t \hat\Lambda)$.
By using the following identity~\cite{wilcox}
\begin{align}
 \frac{\partial}{\partial t} e^{-\beta  \hat{\Lambda}}=-\int_{0}^{\beta} e^{-(\beta-u)  \hat{\Lambda}} \frac{\partial  \hat{\Lambda}}{\partial t} e^{-u  \hat{\Lambda}} \dd u,   
\end{align}
and a Baker-Campbell-Hausdorff formula we find
\begin{align}
    e^{-\frac{i}{\hbar} e \hat\Lambda^{(1)}}\pdv{\hat\Lambda^{(1)}}{t} e^{\frac{i}{\hbar} e \hat\Lambda^{(1)}}=\pdv{\hat\Lambda^{(1)}}{t}-\frac{\ii e}{2\hbar}  \left[\hat\Lambda^{(1)},\pdv{\hat\Lambda^{(1)}}{t}\right].\label{bch}
\end{align}
Note that the second term on the right-hand side is a multiple of the identity for the field Hilbert space, and since it only depends on the position operators (and not the momenta) the higher order BCH terms in \eqref{bch} cancel exactly.

Using $\hat{\bm E}=-\partial_t \hat{\bm A}$, we have
\begin{align}
  \pdv{\hat\Lambda^{(1)}}{t}=& -\int_{\mathbb{R}^3}\dd^3 \bm  R\,\dyad{\bm R}{\bm R}\left[ \hat{\bm r}\cdot \hat{\bm E}(t,\bm R)\right.\nonumber\\
  &\quad\left.+ \frac{\Delta m}{2M}(\hat{\bm r} \cdot \bm\nabla_{\bm R})\left(\hat{\bm r}\cdot\hat{\bm E}(t,\bm R)\right)\right].
\end{align}

There are only two non-vanishing contributions to the commutator of Eq.~\eqref{bch} coming from \eqref{comm 4} and \eqref{comm 7}:
\begin{widetext}
\begin{align}
        \left[\hat\Lambda^{(1)},\pdv{\hat\Lambda^{(1)}}{t}\right]=&\!\int_{\mathbb{R}^3} \!\dd^3 \bm  R \!\int_{\mathbb{R}^3}\! \dd^3 \bm  r \,r_i r_j \!\left(\!\left[ 	\hat{A}^i(t,\bm R),	\partial_{t} \hat{A}^j(t,\bm R)\right]\!+\!\left(\frac{\Delta m}{2M}\right)^2 \!\! r^l r^m \left[ \partial_l	\hat{A}^i(t,\bm R),	\partial_{t}\partial_m \hat{A}^j(t,\bm R)\right]\!\right)\!\dyad{\bm R}{\bm R}\otimes\dyad{\bm r}{\bm r}\nonumber\\
    =&\frac{\ii \hbar}{\epsilon_0}\int_{\mathbb{R}^3} \dd^3 \bm  R \int_{\mathbb{R}^3} \dd^3 \bm  r  \,r_i r_j\left( \delta_{ij}^{(\text{tr})}(0)-\left(\frac{\Delta m}{2M}\right)^2  (\bm r\cdot\bm\nabla_{\bm R})(\bm r\cdot\bm\nabla_{\bm R'}) \left.\delta_{ij}^{(\text{tr})}(\bm R-\bm R')\right|_{\bm R=\bm R'}\right)\dyad{\bm R}{\bm R}\otimes\dyad{\bm r}{\bm r}\nonumber\\
    =&\frac{\ii \hbar}{3\pi^2\epsilon_0}\int_0^{|\bm k_\textsc{uv}|} \dd|\bm k|\, |\bm k|^2\int_{\mathbb{R}^3} \dd^3 \bm  r\, |\bm r|^2\left(1+\frac{1}{5}\left(\frac{\Delta m}{2M}\right)^2|\bm r|^2\right)\dyad{\bm r}{\bm r} \otimes\openone_{\text{COM}}\eqqcolon2\ii \hbar\hat\Delta,\label{divergence}  
\end{align}
\end{widetext}
 where, again, we have a UV cutoff $|\bm k_\textsc{uv}|$ as in the initial dipole expansion of the field.  Eq.~\eqref{divergence} corresponds to Coulombic self-energies which have to be regularized by a cutoff since we initially assumed point charges constituting the atom. They are relevant for Lamb-like energy shifts~\cite{lambshift}.

Combining Eq.~\eqref{cm momentum}, \eqref{rel momentum} and \eqref{bch}, we have now an expression for the transformed Hamiltonian Eq.~\eqref{new H}:
 \begin{align}
   \hat{\tilde{H}}^{(1)}&=\!\!\int_{\mathbb{R}^3} \!\!\!\!\dd^3 \bm  R\! \int_{\mathbb{R}^3} \!\!\!\!\dd^3 \bm  r\dyad{\bm R}{\bm R}\otimes\dyad{\bm r}{\bm r}\!\Bigg(\!\frac{[\hat{\bm{P}}+e\bm{r} \times \hat{\bm{B}}(t,\bm{R})]^{2}}{2 M}\nonumber\\
   &+\frac{\left[\hat{\bm{p}}+\frac{e}{2} \frac{\Delta m}{M}\bm{r} \times \hat{\bm{B}}(t,\bm{R})\right]^{2}}{2 \mu}-\frac{e^2}{4 \pi\epsilon_0 |\bm{r}|} + e^2\hat\Delta\nonumber\\
   &-e \bm{r} \cdot \hat{\bm{E}}(t,\bm{R})-\frac{e}{2} \frac{\Delta m}{M}\left(\bm{r} \cdot \bm{\nabla}_{\bm{R}}\right)[\bm{r} \cdot \hat{\bm{E}}(t,\bm{R})]\Bigg)\label{multi H}.
 \end{align}
As we will be working in the weak-coupling limit, let us discuss and order the terms of Eq.~\eqref{H decompose} according to the two physically relevant small parameters: 1) the coupling strength $e$, and 2) the length-scale of internal state atomic localization in terms of the Bohr radius $a_0$. The latter appears (as we will show later for the leading order terms) through the vanishing of the atomic wavefunctions  for distances from the COM much longer than the Bohr radius. 

Expanding the squares, we can then write Eq.~\eqref{multi H} as a sum of terms with different powers of the small parameters of the problem:
 \begin{align}
      \hat{\tilde{H}}^{(1)}&=\frac{\hat{\bm{P}}^2}{2 M}+\underbrace{\frac{\hat{\bm{p}}^2}{2 \mu}-\frac{e^2}{4 \pi\epsilon_0 |\hat{\bm{r}}|}}_{\text{Hydrogen Hamiltonian}}\nonumber\\*
    &-\underbrace{e \hat{\bm{r}} \cdot \hat{\bm{E}}(t,\hat{\bm{R}})}_{\substack{\text{Electric dipole} \\  \text{$\mathcal{O}(ea_0)$} }}+\underbrace{e\left\{\frac{\hat{\bm{P}}}{2 M},\hat{\bm{r}} \times \hat{\bm{B}}(t,\hat{\bm{R}})\right\}_+}_{\substack{\text{COM R\"ontgen term}\\ \text{$\mathcal{O}(e a_0)$}}}\nonumber\\
    &+\underbrace{e\left\{\frac{\hat{\bm{p}}}{2 \mu},\hat{\bm{r}} \times \hat{\bm{B}}(t,\hat{\bm{R}})\right\}_+}_{\substack{\text{Relative R\"ontgen term (Orbital magnetic dipole)}\\ \text{$\mathcal{O}(e a^2_0)$}}}\nonumber\\
    &-\underbrace{\frac{e\Delta m}{2M}\int_{\mathbb{R}^3} \dd^3 \bm  R \dyad{\bm R}{\bm R}\left(\hat{\bm{r}} \cdot \bm{\nabla}_{\bm{R}}\right)(\hat{\bm{r}} \cdot \hat{\bm{E}}(t,\bm{R}))}_{\substack{\text{Electric quadrupole}\\\text{ $\mathcal{O}(e a^2_0)$}}}\nonumber\\
   &+\underbrace{e^2\hat\Delta}_{\substack{\text{Self-energy}\\\text{ $\mathcal{O}(e^2 a_0^2)$}}}+\underbrace{\frac{e^2}{8\mu}(\hat{\bm{r}} \times \hat{\bm{B}}(t,\hat{\bm{R}}))^2}_{\substack{\text{Diamagnetic term}\\\text{ $\mathcal{O}(e^2 a_0^2)$}}}\nonumber\\
      &\eqqcolon\hat H_0 +\hat H_I+\hat H_{M1}+\hat H_{E2}+\hat H_{\text{dia}}+\hat H_{\text{self}}.\label{H decompose}
 \end{align}
 Let us analyze the different terms one by one. First, we have the unperturbed free atomic Hamiltonian $\hat H_0$ (where the solutions of the relative degrees of freedom are the hydrogenic wavefunctions $\psi_{nlm}(\bm r)$ with an effective mass $\mu$, i.e. the reduced mass) in the form
\begin{align}
\hat{H}_0&=\frac{\hat{\bm{P}}^2}{2 M}+\frac{\hat{\bm{p}}^2}{2 \mu}-\frac{1}{4 \pi\epsilon_0 } \frac{e^{2}}{|\hat{\bm{r}}|}.
\end{align}
 To leading order  $\mathcal{O}(e a_0)$ we then find the electric dipole interaction and the R\"ontgen term associated with the COM motion:
\begin{align}
    \hat{H}_I&=-e \bm{\hat{r}} \cdot \hat{\bm{E}}(t,\hat{\bm{R}})+e\left\{\frac{\hat{\bm{P}}}{2 M},\hat{\bm{r}} \times \hat{\bm{B}}(t,\hat{\bm{R}})\right\}_+.\label{nearly dipole}
\end{align}
 The terms of order $\mathcal{O}(e a^2_0)$ are 1) the electric quadrupole interaction and 2) a \textit{R\"ontgen term} associated with the currents induced by the internal atomic motion, which results in a magnetic dipole coupling with orbital angular momentum degrees of freedom:
\begin{align}
\hat H_{M1}&= e\left\{\frac{\hat{\bm{p}}}{2 \mu},\hat{\bm{r}} \times \hat{\bm{B}}(t,\hat{\bm{R}})\right\}_+,\\
\hat H_{E2}&=-\frac{e\Delta m}{2M}\int_{\mathbb{R}^3} \dd^3 \bm  R \dyad{\bm R}{\bm R}\left(\hat{\bm{r}} \cdot \bm{\nabla}_{\bm{R}}\right)(\hat{\bm{r}} \cdot \hat{\bm{E}}(t,\bm{R})),
\end{align}
The highest order terms in~\eqref{H decompose} with respect to the small parameters are of order  $\mathcal{O}(e^2 a_0^2)$. These are commonly called the diamagnetic and self-energy contributions, respectively:
\begin{align}
  \hat H_{\text{dia}}=& \frac{e^2}{8\mu}(\hat{\bm{r}} \times \hat{\bm{B}}(t,\hat{\bm{R}}))^2,\\
  \hat H_{\text{self}}=&e^2\hat\Delta.
\end{align}

The combined Hamiltonian at leading order in the small parameters  is thus
\begin{align}
  \hat{\tilde{H}}^{(1)}=\hat H_0  +\hat H_I+\mathcal{O}(e^2).\label{general H}
\end{align}
This is the Hamiltonian that we will be studying from here onwards. We will now express  the interaction Hamiltonian $\hat H_I$ in terms of the  hydrogen wavefunctions and COM momentum eigenstates and in the interaction picture generated by $\hat H_0$. Eq. \eqref{nearly dipole} can be rewritten as
 \begin{align}
  \hat{H}_I&=-e \bm{\hat{r}} \cdot \left[ \hat{\bm{E}}(t,\hat{\bm{R}})+\frac{\hat{\bm{P}} \times \hat{\bm{B}}(t,\hat{\bm{R}})- \hat{\bm{B}}(t,\hat{\bm{R}})\times \hat{\bm{P}}}{2M}  \right].\label{rewritten}
 \end{align}
Eq.~\eqref{rewritten} has a very similar structure to Eq.~\eqref{eff dip}. Thus we can (equivalently to the derivation of Sec.~\ref{sec effective dipole}) take the position representation on the relative coordinate by inserting the identity in terms of $\hat{\bm r}$ generalized eigenstates,  and write the atomic dipole operator in the interaction picture of the relative degrees of freedom as 
\begin{equation}
\hat{\bm{d}}\coloneqq   e \bm{\hat{r}} =\sum_{\bm{a}\geq \bm{b}}\hat{\bm{d}}_{\bm{a}\bm{b}}(t,\bm r)
\end{equation}
where the partial dipole between two hydrogenic internal levels $\ket{\Psi_{\bm{a}}}$ and $\ket{\Psi_{\bm{b}}}$ of quantum numbers $\bm a$ and $\bm b$ is
\begin{align}
     \hat{\bm{d}}_{\bm{a}\bm{b}}(t,\bm r)=e \bm F_{\bm{a}\bm{b}}(\bm r) e^{\ii \Omega_{\bm{a}\bm{b}} \tau}\dyad{\bm{a}}{\bm{b}} + \text{H.c.}\label{dip op}
\end{align}
 The spatial smearing vector is given by \mbox{$\bm F_{\bm{a}\bm{b}}(\bm r)=\bm r\Psi_{\bm{a}}^*(\bm r)\Psi_{\bm{b}}(\bm r)$}, and $\hbar\Omega_{\bm{a}\bm{b}}=E_{\bm a}-E_{\bm b}$ is the energy difference between the states $\ket{\bm{a}}$ and $\ket{\bm{b}}$. In contrast to the effective model in Sec.~\ref{sec effective dipole}, the wavefunctions are associated with the reduced mass $\mu$ instead of the electron mass. 
 
 Notice that while the electric dipole in~\eqref{dip op} is smeared with the internal hydrogenic orbitals, the localization of the interaction is not given by these wavefunctions, unlike in the effective model in Eq.~\eqref{eff pos}. Indeed, if we were to evaluate expectations of $\hat H_I$ on a given state of the system, it is the COM localization (the initial state of the COM as an distribution of $\hat{\bm R}$ generalized eigenstates) what gives the spatial localization of the interaction with the field. Of course, the spread of this localization will be bounded from below by the atomic orbital wavefunctions support, but we find that it is the center of mass localization what gives the spatial extension to the atom in the dipole approximation.

It is convenient to take a momentum representation for the COM degrees of freedom in \eqref{rewritten}. We note that for all COM states $\ket{\Psi_\textsc{com}}$
  \begin{align}
     \bra{\bm P}e^{\pm\ii\bm k\cdot\hat{\bm R}}\ket{\Psi_\textsc{com}}= \braket{\bm P\mp\bm k}{\Psi_\textsc{com}}, 
  \end{align}
  and we can identify thus
  \begin{align}
  \bra{\bm P}e^{\pm\ii\bm k\cdot\hat{\bm R}}=\bra{\bm P\mp\bm k}.
\end{align}
Also, we make use of
	\begin{align}
	\hat{\bm E}(t,\bm x)&=\sum_{s=1}^2\! \int_{\mathbb{R}^3} \!\frac{\dd^3 \bm k}{(2 \pi)^{\frac{3}{2}}} \sqrt{\frac{\hbar c |\bm k|}{2 \epsilon_0}} \left( \ii \hat{a}_{\bm k, s} \bm \epsilon_{\bm k, s} e^{\ii  \mathsf{k} \cdot  \mathsf{x}} + \text{H.c.} \right)\label{e field},\\
	\hat{\bm B}(t,\bm x)&=\!\sum_{s=1}^2 \!\int_{\mathbb{R}^3} \!\!\frac{\!\dd^3 \bm k}{(2 \pi)^{\frac{3}{2}}} \sqrt{\frac{\hbar  |\bm k|}{2 c\epsilon_0}} \!\left( \ii \hat{a}_{\bm k, s} (\bm e_{\bm{k}}\cross\bm \epsilon_{\bm k, s}) e^{\ii  \mathsf{k} \cdot  \mathsf{x}} \right.\nonumber\\
	&\quad\left.+ \text{H.c.} \right)\!,
	\end{align}
	where we recall $\bm e_{\bm k}=\bm k/|\bm k|$ is the normalized wave vector. 
Then, the interaction Hamiltonian \eqref{rewritten} in the full Hilbert space interaction picture is given by
  \begin{align}
 \hat{H}_\mathcal{I}&=- \int_{\mathbb{R}^3}\dd^3 \bm  r\, \hat{\bm{d}}(t,\bm r)\cdot\sum_{s=1}^2\! \int_{\mathbb{R}^3} \!\frac{\dd^3 \bm k}{(2 \pi)^{3/2}} \sqrt{\frac{\hbar c |\bm k|}{2 \epsilon_0}}\int \dd^3 \bm  P\nonumber\\
 &\quad\left[\ii \hat{a}_{\bm k, s} e^{-\ii  c|\bm k| t}\bm\alpha_{\bm k,s,\bm P} \dyad{\bm P(t)}{(\bm P-\bm k)(t)}+\text{H.c.}\right],\label{final Hamiltonian}
\end{align}
where we define through the free COM time evolution
\begin{align}
    \ket{\bm P(t)}= \exp\left(\frac{\ii}{\hbar}t\frac{\bm P^2}{2M}\right)\ket{\bm P},\label{energy value}
\end{align}
and 
\begin{align}
   \bm\alpha_{\bm k,s,\bm P}&\coloneqq \bm \epsilon_{\bm k, s}-\frac{ (\bm e_{\bm{k}}\cross\bm \epsilon_{\bm k, s})\cross(\bm P-\hbar\bm k/2)}{Mc}\nonumber\\*
   &=\bm \epsilon_{\bm k, s}\left[1-\frac{\bm P\cdot\bm e_{\bm{k}}-\hbar|\bm k|/2}{Mc}\right]+\bm e_{\bm{k}}\frac{\bm P\cdot\bm \epsilon_{\bm k, s}}{Mc}.\label{pol coeff}
\end{align}
From this, one can see that there is an effective change of the center-of-mass momentum $\bm P$ by $\hbar \bm k/2$ per every plane-wave `component' of the field expansion, as was also noted in \cite{Barnett2017friction}.

Eq.~\eqref{final Hamiltonian} is the final result of our derivation of the interaction Hamiltonian in the interaction picture. It shows that considering a fully quantized atom, the interaction couples all the degrees of freedom:  the (hydrogenic) relative motion degrees of freedom, the center of mass, and the electromagnetic field. 

For a quantum COM, the wavefunction disperses, so one cannot generally find a frame where the momentum  of the COM is exactly zero since momentum eigenstates are unphysical. The best one can do is cancel its expectation value, but the center of mass of any localized atom will still disperse. Thus it is not possible to neglect the R\"ontgen contribution in those cases where the center of mass is a quantum degree of freedom. This is not a problem if the COM degree of freedom is considered classical, where the  R\"ontgen contribution vanishes in the COM comoving frame. Of course, the terms will emerge even in this case if we describe the system in frames where the atom is in motion.

\subsection{Phenomenological example: transition rates}\label{sec transition rate}
In the following, we will treat the dipolar and R\"ontgen interaction terms as a perturbation of the hydrogenic Hamiltonian, so that we can work with the unperturbed internal atomic wavefunctions as a basis to apply perturbation theory. 

Computing transition rates is something well known and addressed many times before in the literature (see, e.g., \cite{Barnett2017}). We include this result  mainly for illustration and completeness but we also generalize it considering initial states that are not necessarily COM momentum eigenstates (which we argued are unnormalizable and unphsyically delocalized). To our knowledge, this assumption has not commonly been relaxed in previous literature.

Consider initially (at time $t=0$) a state of the whole system  $\ket{i,\varphi,0}\coloneqq \ket{i}\otimes\ket{\varphi}\otimes\ket{0}$, where $\ket{i}$ is an energy eigenstate of the internal atomic dynamics, $\ket{0}$ is the EM vacuum, and we allow for the COM to have an arbitrary momentum  distribution: $\ket{\varphi}=\int \dd^3 \bm  P\, \varphi(\bm P)\ket{\bm P}$.
We wish then to compute the transition probability to a different atomic energy level, that is, to a final state  $\ket{f}$ at time $t_f$, where $\ket{f}$ is an energy eigenstate of the internal atomic dynamics. For that we will need to sum over all possible final states for the field and COM degrees of freedom.

To that end we expand in a Dyson series the time evolution operator to first order:
\begin{align}
    \hat{U}=\openone+\hat{U}^{(1)}+\mathcal{O}\left(e^{2}\right),
\end{align}
where $\hat{U}^{(1)}=-\frac{\mathrm{i}}{\hbar} \int_0^{t_f} \dd t\, \hat{H}_{\mathcal{I}}(t)$. 
The probability $P$ for that process to happen at leading order is then
\begin{align}
    P_{\text{tot}}&=\int_{\mathbb{R}^3} \dd^3 \bm  P_f \sum_{s=1}^2\int_{\mathbb{R}^3} \dd^3 \bm  k\\
    &\;\times\left|\int_{\mathbb{R}^3} \!\!\dd^3 \bm  P\, \varphi(\bm P)\matrixel{ f,\bm P_f,1_{\bm k, s}}{\hat{U^{(1)}}}{ i,\bm P,0}\right|^{2}\!\!+\mathcal{O}\left(e^{4}\right).\nonumber
\end{align}
We make use of the following resolution of the identity in the COM and field Hilbert space, respectively,
\begin{align}
    \openone_{\text{COM}}&=\int_{\mathbb{R}^3} \dd^3 \bm P \dyad{\bm P}{\bm P},\\
    \openone_{f}&=\dyad{0}{0}+\sum_{n=1}^\infty\sum_{s=1}^2\int_{\mathbb{R}^3} \dd^3 \bm  k \dyad{n_{\bm k,s}}{n_{\bm k,s}}+... 
\end{align}
The total probability thus reads,
\begin{align}
   P_{\text{tot}}&= \int_{\mathbb{R}^3} \dd^3 \bm  P\, \varphi(\bm P)\int_{\mathbb{R}^3} \dd^3 \bm  P'\, \varphi^*(\bm P') \nonumber\\
    &\quad\matrixel{i,\bm P',0}{\hat{U}^{(1)\dagger}}{ f}\!\!\matrixel{f}{\hat{U}^{(1)}}{i,\bm P,0}+\mathcal{O}\left(e^{4}\right)\nonumber\\
    &=\frac{e^2}{\hbar^2}\int_{\mathbb{R}^3} \dd^3 \bm  P\int_{\mathbb{R}^3} \dd^3 \bm  P'\,  \varphi(\bm P)\varphi^*(\bm P') \nonumber\\
   &\quad \int_0^{t_f} \dd t\int_0^{t_f} \dd t'e^{\ii(\Omega_{fi}t+\Omega_{if}t')}\int_{\mathbb{R}^3}\! \dd^3 \bm  r\!\int_{\mathbb{R}^3}\! \dd^3 \bm  r'\nonumber\\
   &\quad \int_{\mathbb{R}^3} \dd^3 \bm  Q\!\int_{\mathbb{R}^3}\dd^3 \bm  Q'\!\!\int_{\mathbb{R}^3}\! \frac{\text{d}^3\bm{k}}{(2 \pi)^{3}}  \frac{\hbar c |\bm k|}{2 \epsilon_0}e^{\ii c|\bm k|(t-t')}\nonumber\\
    &\quad\times\braket{\bm P'}{\bm Q'(t)}\braket{(\bm Q'-\bm k)(t')}{(\bm Q-\bm k)(t)}\braket{\bm Q(t)}{\bm P}\nonumber\\
    &\quad\sum_{s=1}^2\sum_{a,b=1}^3  F_{fi}^a(\bm r)F_{if}^b(\bm r') \alpha^a_{\bm k,s,\bm Q}\alpha^b_{\bm k,s,\bm Q'}+\mathcal{O}\left(e^{4}\right).\label{ptot-1}
\end{align}
The inner products in Eq.~\eqref{ptot-1} can be thought of as enforcing momentum conservation deltas that yield, upon integration $\bm P=\bm P'=\bm Q=\bm Q'$. Assume now that we are considering a spontaneous decay, i.e. \mbox{$\Omega\coloneqq\Omega_{if}=-\Omega_{fi}>0$}. 
Hence, substituting \eqref{energy value}, Equation \eqref{ptot-1} becomes 
\begin{align}
     P_{\text{tot}}&=\frac{e^2}{\hbar^2}\frac{\hbar c}{2\epsilon_0} \int_0^{t_f} \dd t\int_0^{t_f} \dd t'e^{-\ii\Omega(t-t')}\nonumber\\
     &\quad\sum_{a,b=1}^3\!\int_{\mathbb{R}^3}\!\! \dd^3 \bm  r\, F_{fi}^a(\bm r)\int_{\mathbb{R}^3} \!\!\dd^3 \bm  r'\, F_{if}^b(\bm r')\int_{\mathbb{R}^3} \!\!\dd^3 \bm  P\,|\varphi(\bm P)|^2 \nonumber\\
    &\quad \int_{\mathbb{R}^3}\frac{\dd^3 \bm  k}{(2\pi)^3}|\bm k| e^{\ii(t-t')\left[\frac{\hbar\bm k^2-2 \bm P\cdot\bm k}{2M}+c|\bm k|\right]}   \sum_{s=1}^2\alpha^a_{\bm k,s,\bm P}\alpha^b_{\bm k,s,\bm P}  \nonumber\\
    &\quad   +\mathcal{O}\left(e^{4}\right),
\end{align}
where the summands $\hbar^2\bm k^2/2M- \hbar\bm P\cdot\bm k/M$  correspond to a recoil and Doppler shift, respectively. As is commonplace in the literature, we can take the  limit $t_f\rightarrow\infty$ if we use (Dirac's) Fermi's golden rule for  the transition rate $\Gamma\coloneqq \lim_{t_{f}\rightarrow \infty}\dv{P}{t_f}$:
\begin{align}
  \Gamma&=\frac{e^2}{8 \pi^2\hbar \epsilon_0}\sum_{a,b=1}^3\int_{\mathbb{R}^3} \dd^3 \bm  r\, F_{fi}^a(\bm r)\int_{\mathbb{R}^3} \dd^3 \bm  r'\, F_{if}^b(\bm r') \nonumber\\*
     &\quad \int_{\mathbb{R}^3} \dd^3 \bm  P\,|\varphi(\bm P)|^2\int_{\mathbb{R}^3}\dd^3 \bm  k\,|\bm k|     \sum_{s=1}^2\alpha^a_{\bm k, s,\bm P}\alpha^b_{\bm k, s,\bm P}\nonumber\\
    &\quad\times\delta\left(\frac{\hbar\bm k^2-2 \bm P\cdot\bm k}{2M}+c|\bm k|-\Omega\right) +\mathcal{O}\left(e^{4}\right),\label{gamma-1}
\end{align}
From now on, we will assume that $\varphi(\bm P)=\varphi(|\bm P|)$ such that we can reach closed forms for the integrals. We further define $P\coloneqq|\bm P|$, $k\coloneqq|\bm k|$, and $z\coloneqq e_{\bm P}\cdot e_{\bm k}$. With the help of \eqref{pol coeff} and \eqref{completeness},  we then recast the sum over polarizations in terms of powers of $k$ and $P$:
\begin{align}
    &\sum_{s=1}^2\alpha^a_{\bm k, s,\bm P}\alpha^b_{\bm k, s,\bm P}=\left(\frac{\hbar k}{2Mc}\right)^2 (\delta^{ab}-  e^a_{\bm k}e^b_{\bm k})\nonumber\\*
    &+\frac{\hbar k}{2Mc}\left[2(\delta^{ab}-  e^a_{\bm k}e^b_{\bm k})-\frac{P}{Mc}(2\delta^{ab}z-e^a_{\bm P} e^b_{\bm k}-e^b_{\bm P} e^a_{\bm k}) \right]\nonumber\\*
    &+(\delta^{ab}-  e^a_{\bm k}e^b_{\bm k})-\frac{P}{Mc}(2\delta^{ab}z-e^a_{\bm P} e^b_{\bm k}-e^b_{\bm P} e^a_{\bm k})\nonumber\\*
    &+\left(\frac{P}{Mc}\right)^2\left[z(\delta^{ab}z-e^a_{\bm P} e^b_{\bm k}-e^b_{\bm P} e^a_{\bm k})+ e^a_{\bm k}e^b_{\bm k}\right].\label{decomp}
    \end{align}

Then in spherical coordinates for $\bm k$ and $\bm P$, with $\Theta_k$ and $\Theta_P$ being the respective solid angles, we re-express the integral over $\bm k$ and the angular part of the integral over $\bm P$ in \eqref{gamma-1} as
\begin{align}
    &\int_{S^2}\dd\Theta_P\int_{S^2} \dd\Theta_k\int_0^\infty \dd k\, k^3 \delta\left(\frac{\hbar k^2-2 Pk z}{2M}+ck-\Omega\right)\nonumber\\
    &\quad\times\sum_{s=1}^2\alpha^a_{\bm k, s,\bm P}\alpha^b_{\bm k, s,\bm P}\nonumber\\
    &=\frac{8 \pi}{3}\delta^{ab}\left\{\frac{\hbar^2 \pi\Omega^5}{M^2c^8}\left[1+7\left(\frac{P}{Mc}\right)^2\right]\right.\nonumber\\
    &\quad+\frac{4\hbar\pi\Omega^4}{Mc^6}\left[1-3\frac{\hbar\Omega}{Mc^2}+\left(\frac{P}{Mc}\right)^2\left(5-28\frac{\hbar\Omega}{Mc^2}\right)\right]\nonumber\\
    &\quad+\frac{\pi\Omega^3}{c^4}\left[4-10\frac{\hbar\Omega}{Mc^2}+21\left(\frac{\hbar\Omega}{Mc^2}\right)^2\right.\nonumber\\
    &\left.\left.\quad\!\!+\frac{2}{3}\left(\frac{P}{Mc}\right)^2\!\!\left(20-21\cdot5 \frac{\hbar\Omega}{Mc^2}+18\cdot 21\left(\frac{\hbar\Omega}{Mc^2}\right)^2\right)\right]\right\}\nonumber\\
    &\quad\!\!-\delta^{ab}\frac{16\pi^2\Omega^3}{9c^4}\left(\frac{P}{Mc}\right)^2\!\!\left[12-50\frac{\hbar\Omega}{Mc^2}+147\left(\frac{\hbar\Omega}{Mc^2}\right)^2\right]\nonumber\\
    &\quad+\mathcal{O}\left(\left(\frac{\hbar\Omega}{Mc^2}\right)^6,\left(\frac{P}{Mc}\right)^4\right)\eqqcolon \frac{32\pi^2 M}{3\hbar^3} \delta^{ab}g(P),\label{f}
\end{align}
where we have implicitly defined the function $g(P)$ in the last step.
To solve the integral over $k$ in~\eqref{f}, upon substitution of~\eqref{decomp},  we used that, for general $a_i$,
\begin{align}
    &\int_0^\infty \dd k\, k^3 \delta\left(\frac{\hbar k^2-2 Pk z}{2M}+ck-\Omega\right)(a_2 k^2 +a_1 k+ a_0)\nonumber\\
    &=\theta\!\left( P z -M c(1-\kappa)\right)\sum_{i=0}^2 \frac{a_i \kappa^{3+i}}{\hbar^{3+i}},
\end{align}
where  $\kappa\coloneqq \sqrt{\left(1-\frac{P z}{Mc}\right)^2+\frac{2 \hbar \Omega }{Mc^2}}$, and we also used that \mbox{$P\ll Mc$}.
Finally, we expanded in powers of $P\ll Mc$ as well as $\hbar\Omega/M c^2$ before performing the angular integrals but after the integral over $k$. To perform the angular integrals we made use of
\begin{align}
    \int_{S^2}\dd\Theta_k  \left( \delta^{ab} -  e^a_{\bm k} e^b_{\bm k}\right)=\frac{8\pi}{3}\delta^{ab}.
\end{align}
The expansion in powers of $P\ll Mc$ is justified since  we are working in the non-relativistic regime. However, it is important to note that relativistic corrections of powers higher or equal to $P/Mc$ are not consistent with the approximation made at the level of Equation~\eqref{general H}, since we already neglected the subleading order terms there. Indeed, these relativistic corrections have to be accompanied by the corresponding corrections to the Hamiltonian in order to be consistent (as we will discuss in more detail in Sec.~\ref{sec rel cor}). We will nevertheless keep the subleading corrections in these expressions to analyze qualitatively the dynamics that they generate, but we need to keep in mind that extra corrections from the Darwin terms (Eq.~\eqref{darwin}) would need to be included as well if we want to get numerically accurate predictions. The expansion in powers of $\hbar\Omega/M c^2$ is justified for hydrogenic atoms since the energy of the transitions is much smaller than the rest mass of the atom.

Note that in the case of vacuum excitation processes, i.e. $\Omega\rightarrow -\Omega$, Equation~\eqref{f} vanishes since the argument of the delta is always positive, as we require $P\ll Mc$ to be consistent with the non-relativistic approximation made. However, the fact that the delta argument could be negative outside non-relativistic approximation hints that when we properly include the Darwin correction it may be possible to get `Cherenkov' excitations even in the infinite time limit, as pointed out in~\cite{stritzelberger}.

Concentrating on the sub-leading order in transition frequencies then yields
\begin{align}
    g(P)=&P_0^2\left(1-\frac{3}{2}\frac{\hbar\Omega}{M c^2}+\frac{2}{3}\left(\frac{P}{Mc}\right)^2\right)\nonumber\\
    &+\mathcal{O}\left(\left(\frac{\hbar\Omega}{Mc^2}\right)^5,\left(\frac{P}{Mc}\right)^4\right),\label{f exp}
\end{align}
where we defined 
\begin{align}
    P_0^2=\left(\frac{\hbar\Omega}{M c^2}\right)^3 M^2 c^2.
\end{align}
Let us analyze what kind of phenomenology the subleading corrections generate when we do not consider eigenstates of the COM momentum as initial states and instead consider a COM with a momentum wavefunction $\varphi(P)$. With these definitions, the transition rate yields to leading order
\begin{align}
    \Gamma&=\frac{e^2\Omega^3}{3\pi\hbar\epsilon_0c^3}|\matrixel{i}{\hat{\bm r}}{f}|^2\\*
    &\quad\times 4\pi\int_0^\infty \dd P\,|\varphi(P)|^2\left(1-\frac{3}{2}\frac{\hbar\Omega}{M c^2}+\frac{2}{3}\left(\frac{P}{Mc}\right)^2\right)\nonumber,
\end{align}
where we used that in terms of the internal atomic degrees of freedom $\sum_{a,b=1}^3  F_{fi}^a(\bm r)F_{if}^b(\bm r') \alpha^a_{\bm k,s,\bm P}\alpha^b_{\bm k,s,\bm P}\longrightarrow \sum_{a=1}^3  F_{fi}^a(\bm r)F_{if}^a(\bm r')$.

Let us specialize now to the case of $\ket{i}=\ket{1s}$, i.e. $(n,l,m)=(1,0,0)$, and $\ket{f}=\ket{2p_z}$, i.e. $(n,l,m)=(2,1,0)$. Hence $\hbar\Omega/( M c^2)\approx10^{-8}$, and \mbox{$P_0\approx10^{-30}$~kg~m/s} such that $P_0/Mc\approx 10^{-12}$.
Therefore, one can check that the expansion \eqref{f exp} is valid for $P\lesssim 10^{11} P_0$ such that $P/Mc\lesssim 0.1$. The internal hydrogenic matrix element yields
\begin{align}
   |\matrixel{1s}{\hat{\bm r}}{2p_z}|^2= \sum_{a=1}^3\left|\int_{\mathbb{R}^3} \dd^3 \bm  r F_{2p_z,1s}^a(\bm r)\right|^2=\frac{2^{15}}{3^{10}}a_0^2.
\end{align}
Let us consider, additionally, an initial momentum distribution for the COM $\varphi(P)=(2\pi \sigma_P^2)^{3/4}\exp(-P^2/4\sigma_P^2)$ such that $\ket{\varphi}$ is $L_2$-normalized to one, $\sigma_P$ being the uncertainty in momentum. To leading order then in the expansion of coupling strength, momentum and transition frequency, we arrive at
\begin{align}
    \Gamma=&\frac{e^2 a_0^2 \Omega^3}{3\pi\epsilon_0 \hbar c^3}  \frac{2^{15}}{3^{10}}\left(    1-\frac{3}{2}\frac{\hbar\Omega}{M c^2}+\frac{2}{3}\left(\frac{\sigma_P}{Mc}\right)^2\right)\nonumber\\
    \eqqcolon&\Gamma_0\left(    1-\frac{3}{2}\frac{\hbar\Omega}{M c^2}+\frac{2}{3}\left(\frac{\sigma_P}{Mc}\right)^2\right),\label{gamma dipole}
\end{align}
where $\Gamma_0\approx 6.27\cdot 10^8 /s$ is the well-known hydrogen transition rate expression with no extra corrections~\cite{standardgamma}. It is straightforward to see that in the limit of an initial eigenstate in the COM momentum, i.e. $\sigma_P\rightarrow 0$, $\Gamma_0$ is still shifted due to the finite transition frequency that originated due to the R\"ontgen term.
The expansion is valid for $\sigma_p\ll M c=h/\lambda$, $\lambda$ being the Compton wavelength of the atom. Of course we recall that the corrections proportional to $(\sigma_P/Mc)^2$ will be accompanied by Darwin corrections at the same order. Note that averaging over all $2p$ states, i.e. $m\in\{-1,0,1\}$, would yield the same rate as given by~\eqref{gamma dipole}.


\section{Leading order relativistic corrections}\label{sec rel cor}
 As discussed at the beginning of Sec.~\ref{sec multipolar}, if we are interested in the leading-order correction for relativistic atomic trajectories, we need to include the Darwin Hamiltonian \eqref{darwin}. We include in this section a brief summary of the discussion in~\cite{barnett2018} about how the leading order relativistic corrections would modify the dynamics. Following the same procedure of quantization and PZW transformation as in Sec.~\ref{sec multipolar}, from the minimal coupling Hamiltonian~\eqref{H start} with the Darwin correction~\eqref{darwin} one would arrive at the Hamiltonian~\cite{barnett2018}
 \begin{align}
     \hat H=&\frac{\hat{\bm{P}}^{2}}{2 M}\!\left[1\!-\!\frac{\hat{\bm{P}}^{2}}{4 M^{2} c^{2}}\!-\!\frac{1}{M c^{2}}\left(\frac{\hat{\bm{p}}}{2 \mu}-\frac{e^{2}}{4 \pi \epsilon_{0} |\hat{\bm r}|}\right)\!\right]\!\!-\!\frac{\big(\hat{\bm{P}} \cdot \hat{\bm{p}}\big)^{2}}{2 M^{2} \mu c^{2}}\nonumber\\
     &+\frac{e^{2}}{4 \pi \varepsilon_{0} |\hat{\bm{r}}|} \frac{(\hat{\bm{P}} \cdot \hat{\bm{r}} /|\hat{\bm{r}}|)^{2}}{2 M^{2} c^{2}}
-\frac{\Delta m}{2 \mu M^{2} c^{2}}\left[\left(\hat{\bm{P}} \cdot \hat{\bm{p}}\right) \frac{\hat{\bm{p}}^{2}}{\mu}\right.\nonumber\\
&\left.-\frac{e^{2}}{8 \pi \varepsilon_{0}|\hat{\bm{r}}|}\left( \hat{\bm{P}} \cdot \hat{\bm{p}}+\frac{1}{|\hat{\bm{r}}|^{2}}(\hat{\bm{P}} \cdot \hat{\bm{r}})\left(\hat{\bm{r}} \cdot \hat{\bm{p}}\right)+\text{H.c.} \right)\right]\nonumber\\*
     &+\hat{H}_{\text{A}}+\hat H_I ,\label{H neu darwin}
 \end{align}
 where $\Delta m=m_p-m_e$, and the free internal atomic Hamiltonian 
 \begin{align}
    &\hat{H}_{\text{A}}(\hat{\bm r},\hat{\bm p})=\frac{\hat{\bm p}^{2}}{2 \mu}\left(1-\frac{m_{e}^{3}+m_{p}^{3}}{M^{3}} \frac{\hat{\bm p}^{2}}{4 \mu^{2} c^{2}}\right)\\
    &\quad-\frac{e^{2}}{4 \pi \varepsilon_{0}}\left[\frac{1}{r}+\frac{1}{2 \mu M c^{2}}\left(\hat{\bm p} \cdot \frac{1}{|\hat{\bm r}|} \hat{\bm p}+\hat{\bm p} \cdot \hat{\bm{r}} \frac{1}{|\hat{\bm r}|^{3}} \hat{\bm{r}} \cdot \hat{\bm p}\right)\right]\nonumber
 \end{align}
 no longer assumes the analytically tractable hydrogenic wavefunctions as solutions but a more complicated form. $\hat H_I$ is given by~\eqref{rewritten}, i.e. the dipolar and R\"ontgen interaction to leading order. Significantly, the cross-coupling between COM and internal degrees of freedom of the atom takes a complicated form. For instance, what was the free COM Hamiltonian in the non-relativistic approximation is replaced by the rather non-trivial terms in Eq.~\eqref{H neu darwin} that now have corrections coming from $\hat{\bm P}^2$ and couples the center of mass to the momentum and position operators of the relative motion.  
 
 It is possible in this case to apply a canonical transformation $\{\hat{\bm R},\hat{\bm r},\hat{\bm p}\}\longrightarrow \{\hat{\bm Q},\hat{\bm q},\hat{\bm{\rho}}\}$ that simplfies the form of the corrected Hamiltonian:
 \begin{align}
     \hat{\bm{R}}&= \hat{\bm{Q}}-\frac{\Delta m}{2 M^{2} c^{2}}\left[\left(\frac{\hat{\bm{\rho}}^{2}}{2 \mu} \hat{\bm{q}}+\text{H.c.}\right)-\frac{e^{2}}{4 \pi \varepsilon_{0}|\hat{\bm{q}}|} \hat{\bm{q}}\right]\nonumber\\
     &\quad-\frac{1}{4 M^{2} c^{2}}[(\hat{\bm{q}} \cdot \hat{\bm{P}}) \hat{\bm{\rho}}+(\hat{\bm{P}} \cdot \hat{\bm{\rho}})
     \hat{\bm{q}}+\text{H.c.}],\\
     \hat{\bm{r}}&=\hat{\bm{q}}-\frac{\Delta m}{2 \mu M^{2} c^{2}}[(\hat{\bm{q}} \cdot \hat{\bm{P}}) \hat{\bm{\rho}}+\text{H.c.}]-\frac{\hat{\bm{q}} \cdot \hat{\bm{P}}}{2 M^{2} c^{2}} \hat{\bm{P}},\\
     \hat{\bm{p}}&=\hat{\bm{\rho}}+\frac{\Delta m}{2 M^{2} c^{2}}\left[\frac{\hat{\bm{\rho}}^{2}}{\mu} \hat{\bm{P}}-\frac{e^{2}}{4 \pi \varepsilon_{0}|\hat{\bm{q}}|}\left( \hat{\bm{P}}-\frac{(\hat{\bm{P}} \cdot \hat{\bm{q}}) \hat{\bm{q}}}{|\hat{\bm{q}}|^{2}}\right)\right]\nonumber\\*
     &\quad+\frac{\hat{\bm{\rho}} \cdot \hat{\bm{P}}}{2 M^{2} c^{2}} \hat{\bm{P}}.
 \end{align}
 
 However, the new variables  $\{\hat{\bm Q},\hat{\bm q},\hat{\bm{\rho}}\}$ mix relative motion and COM degrees of freedom. Whereas, the COM momentum is still associated with $\hat{\bm P}$, the remaining new variables lose their original physical meaning of separating internal and external degrees of freedom. 
 After neglecting terms suppressed by $1/M^4c^4$,
 the form of the Hamiltonian becomes~\cite{barnett2018, osborn}:
 \begin{align}
     \hat{\tilde{H}}&=\frac{\hat{\bm{P}}^{2}}{2 M}\left(1-\frac{\hat{H}_{\text{A}}(\hat{\bm q},\hat{\bm p})}{M c^{2}}\right)+\hat{H}_{\text{A}}(\hat{\bm q},\hat{\bm\rho})+ \hat{H}_I,\\
      \hat{H}_I&=-e \hat{\bm{q}} \cdot \hat{\bm{E}}(t,\hat{\bm{Q}})+e\left\{\frac{\hat{\bm{P}}}{2 M},\hat{\bm{q}} \times \hat{\bm{B}}(t,\hat{\bm{Q}})\right\}_+.
 \end{align}
 where the interaction is carried over now in terms of the new canonical variables and the coupling between COM and internal atomic degrees of freedom is more tractable. Additionally, the COM contribution is no longer quartic in the COM momentum.

  \section{Approximate dipole model with classical center-of-mass motion}\label{sec eff dipole 2}

After having studied the nuances related to taking into account the COM dynamics, one realizes quickly that it would be truly challenging to consider scenarios where the COM trajectories undergo arbitrarily accelerated relativistic motion since the coupling of internal and external degrees of freedom becomes increasingly complicated. This poses the question whether we can use effective models that a) allow for the COM motion to be relativistic b) are computationally tractable and c) are reasonable approximations that at least capture the main phenomenology of an interaction between matter and light.

With this in mind, let us come back to the effective dipole model from Sec.~\ref{sec effective dipole}. Now it becomes clear that we are neglecting the quantum nature of the COM and along with it the dynamics in form of the R\"ontgen term of the COM.
However, in contrast to the multipolar Hamiltonian, we can consider relativistic, and externally prescribed trajectories of the atom. Furthermore, if this model holds any value for the probing of the electromagnetic field, the predictions should be generally covariant for different observers. This is in distinction to the multipolar Hamiltonian that can only guarantee Galilei covariance. One would expect that although this model may not give the accurate numbers associated with a particular atomic physics experiment, it may still hold some of the core phenomenology of the light-matter interaction and provide a simple yet covariant model to measure the electromagnetic field. Neglecting the R\"ontgen term for the COM would be akin to considering that a) the COM is a classical degree of freedom and b) the mass of the nucleus is much larger than that of the electron.

Similar to the PZW transformation \eqref{gauge 2} being applied to the two-particle minimal coupling Hamiltonian \eqref{H start} to find the multipolar Hamiltonian, one can consider the transformation
\begin{align}
    \hat{\mathcal{U}}_1=\exp(\frac{\ii}{\hbar}e\hat{\Lambda}_{1}),
\end{align}
generated by 
\begin{align}
     \hat{\Lambda}_{1}(t,\hat{\bm r}_e)= \hat{\bm r}_e \cdot\int_{0}^{1} \dd u \hat{\bm{A}}\left(t,u \hat{\bm r}_e\right)\label{effective gauge}
\end{align}
applied to the one-particle minimal coupling Hamiltonian 
\begin{align}
\hat H_{\text{eff}}=\frac{1}{2m_e}(\hat{\bm p}_e + e \hat{\bm A}(t,\hat{\bm r}_e))^2 -e \hat U(t,\hat{\bm r}_e)-\frac{e^{2}}{4 \pi\epsilon_0 |\hat{\bm{r}}_e|}
\end{align}
to arrive at the effective dipole model \eqref{eff dip}.
In contrast to the case where there is COM dynamics, \eqref{effective gauge}  is a gauge transformation  where the transformed potentials can be expressed in terms of the electric and magnetic field~\cite{Kobe84}:
\begin{align}
 &\hat{\mathcal{U}}_1\hat{\bm A}(t,\hat{\bm r}_e) \hat{\mathcal{U}}_1^\dagger+\frac{1}{ e}\hat{\mathcal{U}}_1 \hat{\bm p}_e \hat{\mathcal{U}}_1=-\hat{\bm r}_e\times\int_0^1 \dd u~ u \hat{\bm B}(t,u\hat{\bm r}_e), \nonumber\\
  & \hat{\mathcal{U}}_1 \hat{U}(t,\hat{\bm r}_e) \hat{\mathcal{U}}_1^\dagger+\frac{\ii\hbar}{ e}\hat{\mathcal{U}}_1 \partial_{t} \hat{\mathcal{U}}_1=-\hat{\bm r}_e\cdot\int_0^1 \dd u~ \hat{\bm E}(t,u\hat{\bm r}_e)+\hat{\delta},
\end{align}
where $\hat{\delta}$ is a self-energy that needs to be regularized---which introduces corrections of $\mathcal{O}(e)$. In the dipole approximation, to leading order in coupling $e$ and Bohr radius $a_0$, the Hamiltonian then yields
\begin{align}
    \hat H_{\text{eff}}=&\frac{\hat{\bm p}^2_e}{2m_e}-\frac{1}{4 \pi\epsilon_0 } \frac{e^{2}}{|\hat{\bm{r}}_e|} +e \hat{\bm r}_e\cdot\hat E(t,\hat{\bm r}_e),
\end{align}
which is precisely the effective model of Sec.~\ref{sec effective dipole}.

We recall that this effective model is prescribed in the center-of-mass frame of the atom, where the atom does not move, and hence there are no COM R\"ontgen terms. We denote $\tau$ the proper time of the atom's COM rest frame $(\tau,\bm \xi)$. As common in particle detector models, we take the atom to be Fermi-Walker transported as the interatomic forces preserve its spatial coherence~\cite{tales1}. It is convenient to quantize the field in an inertial frame that we will call the `lab' frame of coordinates $(t,\bm x)$. For general spacetimes characterized by the metric $g$ the interaction Hamiltonian that generates translations with respect to the COM proper time (in the interaction picture) reads
\begin{align}
     \hat H^\tau_{I,\text{eff}}  &=\chi(\tau)\!\sum_{\bm{a}> \bm{b}} \! \int_{\Sigma_\tau}\!\!\!\!\dd^3 \bm  \xi\sqrt{-g}~ \hat{\bm{d}}_{\bm{a}\bm{b}}(\tau,\bm \xi) \cdot\hat{\bm{E}}\left(t(\tau,\bm \xi),\bm x(\tau,\bm \xi)\right) \nonumber\\*
     &=\int_{\Sigma_\tau} \dd^3 \bm \xi\, \hat{h}_{I,\text{eff}}(\tau,\bm\xi),\label{eff prop}
\end{align}
where the time dependence of the coupling is prescribed in the COM rest frame and encoded in $\chi$, and where we defined the Hamiltonian density $\hat{h}_{I,\text{eff}}(\tau,\bm\xi)$.


\subsection{Covariance of predictions}

If the model holds any value as a relativistic probe of the electromagnetic field, its predictions in flat spacetime should be Lorentz covariant. To show explicitly that they are, we take  \eqref{eff prop} and analyze how the Hamiltonian transforms under changes of reference frame.  For Minkowski spacetime in any coordinates associated with internal frames we have that $\sqrt{-g}=1$. Assuming that the atom is undergoing inertial motion, we can compute the Hamiltonian that generates translations with respect to the lab frame using  the transformation properties under general Lorentz transformations. 
 The covariance of the model demands that
   \begin{align}
       \hat U&=\mathcal{T} \exp \left(\frac{-\mathrm{i}}{\hbar} \int_{\mathbb{R}^3\times \mathbb{R}}\!\!\!\!\!\!\!\!\! \mathrm{d}^{3} \bm\xi\, \mathrm{d} \tau\, \hat{h}_{I,\text{eff}}(\tau,\bm\xi)\right)\nonumber\\
&=\mathcal{T} \exp \left(\frac{-\mathrm{i}}{\hbar} \int_{\mathbb{R}^3\times \mathbb{R}}\!\!\!\!\!\!\!\!\! \mathrm{d}^{3}\, \boldsymbol{x}\, \mathrm{d} t\, \hat{h}_{I,\text{eff}}(\tau(t,\bm x),\bm \xi(t,\bm x))\right).
 \end{align}
In the Hamiltonian \eqref{eff prop}, the electric field is as seen from the COM frame. However, it is quantized in the lab frame. To write the Hamiltonian that generates translations with respect to the lab frame's time $t$ we need to transform the electric field. Let us assume then that the atomic COM moves on a trajectory $\bm x(t)=vt$ and velocity $v$ with respect to the lab frame.
 The electric field is a spatial vector under Lorentz transformations:
\begin{align}
    &\hat{\bm{E}}(t(\tau,\bm\xi),x((\tau,\bm\xi))\rightarrow\\&\gamma\left(\hat{\bm{E}}(t,\bm x)+\bm v \cross \hat{\bm{B}}(t,\bm x)\right) + (1-\gamma) \left(\hat{\bm{E}}(t,\bm x) \cdot \bm e_v\right) \bm e_v,\nonumber\label{e transform}
\end{align}
where $\bm e_v=\bm v/|\bm v|$.
The Lorentz transformed Hamiltonian  generating translations with respect to time $t$ is thus
\begin{align}
    \hat H^t_{I,\text{eff}}&= \sum_{\bm{a}>\bm{b}} \int_{\mathbb{R}^3}\dd^3 \bm  x\,  \chi(\tau(t,\bm x)) \hat{\bm{d}}'_{\bm{a}\bm{b}}(\tau(t,\bm x),\bm\xi(t,\bm x))\nonumber\\
    &\quad\times  \left\{\gamma[\hat{\bm{E}}(t,\bm x)+\bm v \cross \hat{\bm{B}}(t,\bm x)]\right.\nonumber\\
    &\quad\quad\quad\left.+ (1-\gamma) \left(\hat{\bm{E}}(t,\bm x) \cdot \bm e_v\right) \bm e_v\right\}.
\end{align}
Naturally, a R\"ontgen term arises for the classical COM through the Lorentz transformation.
The transformed dipole moment reads 
\begin{align}
\hat{\bm{d}}'_{\bm{a}\bm{b}}(\tau(t,\bm x),\bm\xi(t,\bm x))&=\!e\bm F_{ab}(\xi(t,\bm x)) e^{\ii \Omega_{\bm{a}\bm{b}} \tau(t,\bm x)}\!\dyad{\bm{a}}{\bm{b}} + \text{H.c.}   
\end{align}

Although it is not necessary to prove that this is covariant because it was made covariant by construction, there is some value in explicitly showing its covariance and how to deal with changes of reference frame in the context of this effective light-matter interaction. With this in mind let us compute the transition probability of the atom in the COM frame and the lab frame explicitly showing how they coincide.

\subsection{Example - Vacuum excitation probability}

We will showcase a simple example to demonstrate that the previous considerations yield Lorentz invariant predictions. Let us consider an atom whose COM is comoving with the lab frame. Let us compute the transition probability from a $\ket{1s}$ state to the excited state $\ket{2p_z}$. Let us do this calculation using two different coordinate systems. One comoving with the atomic COM and the lab frame and another one moving at a constant speed with respect to the lab frame, showing how both results coincide. This will also allow us to compute  very useful quantities along the way such as the Wightman tensor for the electric and magnetic fields. 

\subsubsection{Wightman functions}

We will give first the electromagnetic Wightman functions which will be used to compute the subsequent transition probabilities (the derivations can be found in Appendix~\ref{app-wightmann}):
 We begin with the  electric field two-point function which is of the form
\begin{align}
   & W^{ij}_E[t,t';\bm x,\bm x']= \bra 0 \hat{E}^i(t,\bm x) \hat{E}^j(t',\bm x') \ket 0\nonumber\\
   &=\frac{\hbar c}{2\epsilon_0} \int_{\mathbb{R}^3}  \frac{\dd^3 \bm  k}{(2 \pi)^{3}} |\bm k| e^{-\ii c|\bm k| (t-t')} e^{\ii \bm k \cdot (\bm x - \bm x')} \left( \delta^{ij} -  e^i_{\bm k} e^j_{\bm k}\right),\label{WE 1}
\end{align}
and can be put in relation to the magnetic field Wightman tensor
\begin{align}
    W^{ij}_B[t,t';\bm x,\bm x']&=\bra 0 \hat{B}^i(t,\bm x) \hat{B}^j(t',\bm x') \ket 0\nonumber\\
   &=\frac{1}{c^2}W^{ij}_E[t,t';\bm x,\bm x'] \label{WB2}.
\end{align}
The two cross-field Wightman functions can be similarly related:
\begin{align}
& W^{ij}_{BE}[t,t';\bm x,\bm x']=\bra 0 \hat{B}^i(t,\bm x) \hat{E}^j(t',\bm x') \ket 0\nonumber\\*
  &=-\frac{\hbar }{2\epsilon_0} \int_{\mathbb{R}^3}  \frac{\dd^3 \bm  k}{(2 \pi)^{3}} |\bm k| e^{-\ii c|\bm k| (t-t')} e^{\ii \bm k \cdot (\bm x - \bm x')}     \epsilon^{ijk} (e_{\bm k})_k\nonumber\\*
  &=W^{ji}_{EB}[t,t';\bm x,\bm x']\label{WEB3}.
  \end{align}
while all the details can be seen in Appendix~\ref{app-wightmann}.
We additionally give an explicit form for the different electromagnetic Wightman functions after performing the integral over $\bm k$. The following expansion in terms of spherical harmonics $Y_{lm}$ and spherical Bessel functions of the first kind $j_l$~\cite{NIST:DLMF} will be of use:
 \begin{align}
 	&	\bm k= \sqrt{\frac{2\pi}{3}}|\bm k|\Big( Y_{1 1}(\bm e_{\bm k})- Y_{1 -1}(\bm e_{\bm k}),\ii [Y_{1 1}(\bm e_{\bm k})- Y_{1 -1}(\bm e_{\bm k})]\!,\nonumber\\*
 	&\left.\quad\quad\quad\quad\sqrt{2 }Y_{1 0}(\bm e_{\bm k})\right),\label{wave}\\*
 	&	e^{\ii \bm k \cdot \bm x}= \sum_{l=0}^\infty \sum_{m=-l}^l 4 \pi \ii^l j_l(|\bm k| |\bm x|) Y_{l m}(\bm e_{\bm k})Y^*_{l m}(\bm e_{\bm x})\nonumber\\*
 	&\quad\quad=\sum_{l=0}^\infty \sum_{m=-l}^l 4 \pi \ii^l j_l(|\bm k| |\bm x|) Y^*_{l m}(\bm e_{\bm k})Y_{l m}(\bm e_{\bm x}).\label{plane}
\end{align}
Then, the two independent Wightman functions  $W_E$ and $W_{BE}$  (with implicit pole prescription, and derivative $'$ with respect to $\tilde{r}$) are: 
\begin{align}
     &W^{ij}_E[t,t';\bm x,\bm x']=\frac{\hbar  c}{2(2\pi)^2\epsilon_0}\\*
     &\times\left[\frac{8  [\tilde{r}^2 (2 X^{ij}+\delta^{ij})- c^2(t'-t)^2 \delta^{ij}] }{\left(\tilde{r}^{2}-c^2(t'-t)^{2}\right)^{3}} \right.\nonumber\\*
     &\quad+ \ii\pi\left\{\frac{\delta^{\prime \prime}(\tilde{r}-c(t'-t))-\delta^{\prime \prime}(\tilde{r}+c(t'-t)) }{\tilde{r}} X^{ij}\right.\nonumber\\*
    &\quad\quad+(3 X^{ij} +2 \delta^{ij} )\left( \frac{\delta(\tilde{r}-c(t'-t))-\delta(\tilde{r}+c(t'-t))}{\tilde{r}^3}\right.\nonumber\\*
    &\quad\quad\quad\left.\left.\left.-\frac{ \delta^{\prime}(\tilde{r}-c(t'-t))- \delta^{\prime}(\tilde{r}+c(t'-t))}{\tilde{r}^2}\right)\right\}\right],\nonumber\\*
         &W^{ij}_{BE}[t,t';\bm x,\bm x']=-\frac{\hbar}{2 (2\pi)^2 \epsilon_0} \beta \epsilon^{ijk}(e_{\bm x}-e_{\bm x'})_k,
\end{align}
where we defined $|\bm x-\bm x'|=\tilde{r}$, $e_{\bm x}=\bm x/|\bm x|$, and
\begin{align}
    X^{ij}&=(e_{\bm x}-e_{\bm x'})^i(e_{\bm x}-e_{\bm x'})^j- \delta^{ij},\\
    \beta&=\frac{16    c(t'-t) \tilde{r}}{\left(\tilde{r}^{2}-c^2(t'-t)^{2}\right)^{3}}\\
    &+\ii \pi\left[\frac{\delta^{\prime \prime}(\tilde{r}-c(t'-t))+\delta^{\prime \prime}(\tilde{r}+c(t'-t))}{\tilde{r}^2}\right.\nonumber\\*
    &\quad\left.- \frac{\delta^{'}(\tilde{r}-c(t'-t))+ \delta^{'}(\tilde{r}+c(t'-t))}{\tilde{r}^3}\right]\nonumber.  
\end{align}
These results can be confirmed for the real part in chapter 9 of \cite{takagi} and for the imaginary part~\cite{cohen} (by noting that that the imaginary part of the Wightman corresponds  to the commutators of the respective fields).


\subsubsection{Calculation in the COM/lab frame}\label{sec lab frame}
 Let us first calculate the  transition probability assuming that the atom is at rest in the lab frame. Without loss of generality we can assume that rest and lab frame are identical $(t,\bm x)=(\tau,\bm \xi)$. We can then perform a perturbative analysis. We compute the Dyson series of the time evolution operator to first order:
\begin{align}
    \hat{U}=\openone+\hat{U}^{(1)}+\mathcal{O}\left(e^{2}\right),
\end{align}
where $\hat{U}^{(1)}=-\frac{\mathrm{i}}{\hbar} \int_{-\infty}^{\infty} \mathrm{d} t \hat{H}_{I,\text{eff}}(t)$. The vacuum excitation probability  for the initial joint ground state reads 
\begin{align}
    \mathcal{P}&=\SumInt_{\text{out}}|\matrixel{ 2p_z, \text {out}}{\hat{U}}{ 1s, 0}|^{2}\nonumber\\
    &=  \!\SumInt_{\text { out }}\!\!\matrixel{1s, 0}{\hat{U}^{(1)\dagger}}{ 2p_z, \text{out}}\!\!\matrixel{ 2p_z, \text{out}}{\hat{U}^{(1)}}{ 1s, 0}\!+\mathcal{O}\left(e^{4}\right)\nonumber\\
    &=\frac{e^2}{\hbar^2}\int_\mathbb{R} \dd t\int_\mathbb{R}\dd t' \,\chi(t)\chi(t') e^{\ii\Omega_{2p_z 1s}(t'-t)}\nonumber\\*
    &\times\!\int_{\mathbb{R}^3}\!\!\dd^3 \bm x\int_{\mathbb{R}^3}\dd^3 \!\!\bm x' \,\bm F_{2p_z 1s,i}(\bm x) W_E^{ij}[t,t';\bm x,\bm x'] \bm F_{2p_z 1s,j}(\bm x')  \nonumber\\*
    &+\mathcal{O}\left(e^{4}\right), \label{probability}
\end{align}
where we used the resolution of identity in terms of the field states $\ket{\text{out}}$.  Using Eq.~\eqref{WE 1}, we can write 
\begin{align}
      \mathcal{P}&= \frac{e^2c}{2(2 \pi)^{3}\epsilon_0\hbar} \int_{\mathbb{R}^3} \dd^3 \bm  k\,|\bm k| \left|\int_{\mathbb{R}}\dd t\, \chi(t) e^{-\ii(\Omega_{2p_z 1s}+c|\bm k|)t}\right|^2\nonumber\\
     &\quad\times\left[\left|\int_{\mathbb{R}^3}\dd^3 \bm  x\, e^{\ii \bm k \cdot \bm x } \bm F_{2p_z 1s}(\bm x) \right|^2\right.\\
     &\quad\quad\left.-\left|\int_{\mathbb{R}^3}\dd^3 \bm  x\, e^{\ii \bm k \cdot \bm x } \bm F_{2p_z 1s}(\bm x)\cdot\bm e_{\bm k} \right|^2\right]+\mathcal{O}\left(e^{4}\right)\nonumber.
\end{align}


 For simplicity and also comparison with previous works, we can further assume that the time-dependent coupling is of Gaussian adiabatic nature, i.e. $\chi(t)=\exp(-(t/T)^2)$ with $T$ being the time scale of interaction (A discussion on the physicality of such a switching function for the light-matter interaction can be found in, e.g, \cite{randomness}). After a lengthy but simple calculation that parallels the calculation in Appendix A of ~\cite{randomness}, using~Eq.~\eqref{wave} and \eqref{plane} and the fact that for $f_{lm}\in\mathbb{C}$ it is satisfied that
  \begin{align}
 &\int_{S^2} \dd \Theta_k \left|\sum_{l=0}^\infty\sum_{m=-l}^l f_{l m} Y_{l m}(\bm e_{\bm k})\right|^2=\sum_{l=0}^\infty\sum_{m=-l}^l \left| f_{l m}\right|^2.
\end{align}
 the probability can then be evaluated to
\begin{align}
  \mathcal{P}=&\,49152\frac{(e a_0 T)^2 c}{\pi \hbar\epsilon_0 }     \int_{0}^{\infty}\dd |\bm k|  \frac{|\bm k|^3 e^{-\frac{1}{2} T^2(c |\bm k|+\Omega_{2p_z 1s} )^2}}{\left(4 a_0^2 |\bm k|^2+9\right)^6}\nonumber\\*
  &+\mathcal{O}\left(e^{4}\right).\label{check}
\end{align}
Using natural units for a hydrogen atom (the generalization to an hydrogenoid atom is straightforward) with $c=\hbar=\epsilon_0=1, e\approx 137^{-1/2}$, $a_0\approx 2.68\times 10^{-4}~\text{eV}^{-1}$, $m\approx5.1\times 10^5~\text{eV}$ and $\Omega\approx3.73 ~\text{eV}$, we plot the vacuum excitation probability in Fig.~\ref{plot-stationary} which will be our reference point for the calculations of the next section.

\subsubsection{Calculation for a boosted observer}
Now let us compare the previous result with the transition probability as computed by an observer that moves in the $z$ direction as seen from the lab frame with velocity $v$. This corresponds to a Lorentz boost in the $z$ direction, i.e. $\{c\tau=\cosh{\eta}\,c t-\sinh{\eta}\,x_3,\xi_3=\cosh{\eta}\,x_3-\sinh{\eta}\,c t \}$ with rapidity $\eta$.
As the electric field transforms via \eqref{e transform}, the electric field Wightman tensor in the probability expression of~\eqref{probability} transforms as,
\begin{widetext}
\begin{align}
  & W^{ij}_E[t,t';\bm x,\bm x']\rightarrow \nonumber\\
   &\gamma^2 \left(W^{ij}_E[t,t';\bm x,\bm x']+\epsilon_{abi}\epsilon_{cdj} v^a v^c W^{bd}_B[t,t';\bm x,\bm x']  + \epsilon_{abi} v_a W^{bj}_{BE}[t,t';\bm x,\bm x']+\epsilon_{cdj} v^c W^{id}_{EB}[t,t';\bm x,\bm x'] \right)\nonumber\\
   &+(1-\gamma)^2  (e_{v})_a (e_{v})_b W^{ab}_E[t,t';\bm x,\bm x'] (e_{v})^i(e_{v})^j+ \gamma(1-\gamma)\Big[W^{ia}_E[t,t';\bm x,\bm x'] (e_{v})_a(e_{v})^j+W^{aj}_E[t,t';\bm x,\bm x'] (e_{v})_a(e_{v})^i \nonumber\\
   &\quad+\epsilon_{abi}v^a W^{bc}_{BE}[t,t';\bm x,\bm x']+\epsilon_{cdj}v^c W^{ad}_{EB}[t,t';\bm x,\bm x'](e_{v})_a(e_{v})^i\Big]. 
\end{align}
We can then use $v_a=v\delta_{a3}$, and the relations~\eqref{WE 1}, \eqref{WB2} and \eqref{WEB3} to arrive at
\begin{align}
&W_E[t,t';\bm x,\bm x']\rightarrow\frac{\hbar c}{2\epsilon_0} \int_{\mathbb{R}^3}  \frac{\dd^3 \bm  k}{(2 \pi)^{3}} |\bm k| e^{-\ii c|\bm k| (t-t')} e^{\ii \bm k \cdot (\bm x - \bm x')} M,
\end{align}
where \begin{align}
 M=   \left(
\begin{array}{ccc}
\gamma^2\left(1- e^3_{\bm k} \frac{v}{c} \right)^2- e^1_{\bm k}e^1_{\bm k} &  -e^1_{\bm k}  e^2_{\bm k} &\gamma e^1_{\bm k} \left(\frac{v}{c}- e^3_{\bm k}\right) \\
 - e^1_{\bm k}  e^2_{\bm k} & \gamma^2\left(1- e^3_{\bm k} \frac{v}{c}\right)^2-e^2_{\bm k}e^2_{\bm k} & \gamma e^2_{\bm k} \left(\frac{v}{c}-e^3_{\bm k}\right) \\
 \gamma e^1_{\bm k} \left(\frac{v}{c}- e^3_{\bm k}\right) &  \gamma e^2_{\bm k} \left(\frac{v}{c}-e^3_{\bm k}\right) & 1-e^3_{\bm k}e^3_{\bm k} \\
\end{array}
\right).
\end{align}
\end{widetext}

\begin{figure}[!h]
    \centering
    \includegraphics[scale=.57]{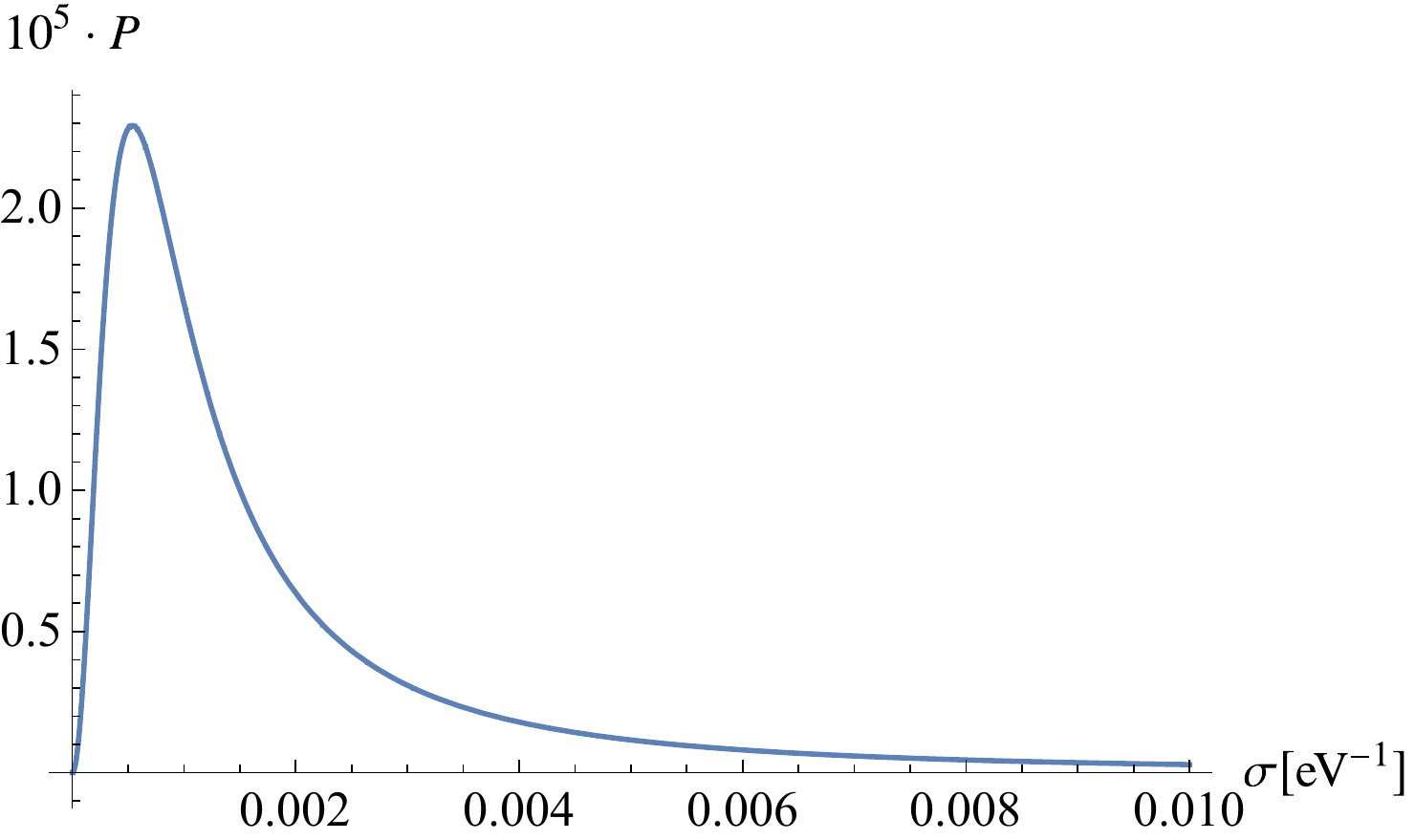}
    \caption{Vacuum excitation probability to the first excited state $2p_z$ for a stationary hydrogen atom and stationary observer as a function of the time scale of the interaction $T$ ($e\approx 137^{-1/2}$, $a_0\approx 2.68\times 10^{-4}~\text{eV}^{-1}$, $m\approx5.1\times 10^5~\text{eV}$ and $\Omega\approx3.73 ~\text{eV}$).}
    \label{plot-stationary}
\end{figure}

The excitation probability therefore  becomes
\begin{align}
     \mathcal{P}=&  \frac{e^2}{\hbar^2}\frac{\hbar c}{2\epsilon_0}\int_{\mathbb{R}^3}  \frac{\dd^3 \bm  k}{(2 \pi)^{3}}|\bm k|\iint_{\mathbb{R}^2} \dd t\dd t' e^{-\ii c|\bm k| (t-t')}\nonumber\\
    &\times\iint_{\mathbb{R}^6}\dd^3 \bm  x\dd^3 \bm  x' e^{\ii \bm k \cdot (\bm x-\bm x')} \nonumber \\
    &\times\chi(\tau(t,\bm x))\chi(\tau(t',\bm x')) e^{\ii\Omega_{2p_z 1s}(\tau(t',\bm x')-\tau(t,\bm x))} \nonumber\\ 
    &\times\bm F_{2p_z 1s}(\bm\xi(t,\bm x)) \cdot M \cdot \bm F_{2p_z 1s}(\bm\xi(t',\bm x'))   +\mathcal{O}\left(e^{4}\right).\label{transformed prob}
\end{align}
With the change of variables (applied twice) $\{ct=\cosh{\eta}\,c \tau+\sinh{\eta}\,\xi_3,x_3=\cosh{\eta}\,\xi_3+\sinh{\eta}\,c \tau \}$,  which is equivalent to the inverse Lorentz transformation (and thus non-singular), we get exactly  the same result as in the proper frame of the atom for the transition probability in Eq.~\eqref{check}. Similarly one can check numerically that Eq.~\eqref{transformed prob} reproduces Fig.~\ref{plot-stationary}. It is clear then that, since the model is covariant, choosing a convenient frame, the atomic rest frame in this example, significantly simplifies the calculations.

\section{Implications for the Unruh-DeWitt model}\label{sec modify udw}

One of the most common scalar approximations to the light-matter interaction is the use of the UDW model presented in Section~\ref{sec effective UDW}. This model can certainly approximate the light matter interaction under the effective dipole approximation when we consider a `heavy' atom with a classical center of mass even when the atomic motion is relativistic as discussed in a number of previous papers~\cite{Martin-MartinezMOnteroDelRey,Pablo1,pozas}. 

For the effective dipole model, we can always describe the interaction in the comoving frame of the atom where there would be no R\"ontgen term, and the corresponding R\"ontgen terms in other frames emerge out of the reference frame transformations as described in Sec.~\ref{sec eff dipole 2}. However, after the analysis of the dynamics of the atomic center of mass  and the internal degrees of freedom of the atom, one may wonder whether the usual scalar approximations to the light-matter interaction (such as the Unruh-DeWitt model) can be `upgraded' to phenomenologically capture (still with a simple scalar model) the effect of the missing  R\"ontgen terms outside of the `infinitely heavy' proton approximation of the effective dipole model. This is particularly relevant when one has a quantum center of mass which is necessarily delocalized in momentum since momentum eigenstates are nonphysical.

Based on the interaction Hamiltonian~\eqref{nearly dipole} we propose the following scalar analogue of the dipole interaction:
\begin{align}
    \hat H_{\text{Scalar}}= \hat H_{\text{Mono}}+\hat H_{\text{R\"o}}.
\end{align}
That is, the sum of a monopole moment like in the UDW model and an analogue scalar R\"ontgen term. This monopole term has the peculiarity that spatial localization is given in terms of the COM wavefunction (as was also argued in~\cite{stritzelberger}). The new monopole interaction then reads
\begin{align}
    \hat H_{\text{mono}}=\lambda\hat{\mu}\otimes\int_{\mathbb{R}^3} \dd^3 \bm  R~\hat\phi(\bm R)\dyad{\bm R}{\bm R}.
\end{align}
We also need to add an effective coupling of the internal, COM, and field degrees of freedom mimicking the R\"ontgen interaction of~\eqref{nearly dipole}. 
This interaction is vectorial in its core, so it is very difficult to capture its behaviour in an analog scalar model. As we will see, a qubit UDW detector is not naturally well-suited to build such an analogy outside the 1+1-dimensional case.  Further, we need an analogue of the magnetic field operator to build up our R\"ontgen facsimile. 

Our analogy starts between the $(n+1)$-dimensional scalar field as an expansion in plane wave modes, 
\begin{align}
    \hat{\phi}(t,\bm x)=\int \frac{\mathrm{d}^{n} \bm{k}}{(2 \pi)^{n / 2} \sqrt{2 \omega}}\left(e^{-\mathrm{i} \omega  t+\mathrm{i} \bm{k} \cdot \bm{x}} \hat{a}_{\boldsymbol{k}}+\text{H.c.}\right)
\end{align}
and the electric field~\eqref{e field}. We have also a relation between the magnetic and electric field through the Maxwell equation $\partial_{0} \hat{\bm E}=\bm\nabla\times \hat{\bm B}$ (without external currents). Finding a scalar analogue of this equation that is so remarkably vectorial in nature  will come at the price of some ambiguities and choices in the model. From the Heisenberg equation of motion we know that $\partial_0 \hat \phi=\hat\pi$, where 
\begin{align}
    \hat{\pi}(t, \bm{x})=- \int \frac{\mathrm{d}^{n} \boldsymbol{k}}{(2 \pi)^{n / 2}} \sqrt{\frac{\omega }{2}}\left(\ii e^{-\mathrm{i} \omega  t+\mathrm{i} \bm{k} \cdot \bm{x}} \hat{a}_{\boldsymbol{k}}+\text{H.c.}\right)
\end{align}
is the canonical momentum operator to $\hat\phi$. 
As there is necessarily a limitation in the alignment of scalar and vector theory, we suggest here to find an operator $\grad\hat X$ mimicking the magnetic field such that $ \hat\pi=(\grad\hat X)\cdot \bm \epsilon$ is satisfied, for some spatial direction $\bm \epsilon$.

In $(1+1)$D it is straightforward to find the operator
\begin{align}
     \partial_x\hat X(t, \bm{x})=-\partial_x \hat{\phi}(t,\bm x),
\end{align}
being nothing else than the spatial derivative of the field operator itself. 
Therefore, an analogous construction in $(1+1)$ dimensions for the R\"ontgen term of the UDW model would read 
 \begin{align}
  \hat H_{\text{R\"o}}= - \lambda\int_{\mathbb{R}} \dd R\left\{\frac{\hat{P}}{2 M},\hat{\mu}\cdot\partial_R\hat{\phi}(R)\right\}_+\dyad{ R}{ R}.
 \end{align}
To better capture an analogy to the R\"ontgen term in (3+1) dimensions let us therefore model from here on the internal detector degrees of freedom through a quantum harmonic oscillator (as it is also common for UDW detectors~\cite{hu2007, hu2008, Ostapchuk, dragan2011, brown2013}) with respective position and momentum operators $\hat{\bm q}$ and $\hat{\bm p}_q$.
We suggest then that our analogue magnetic field should be $-\grad\hat\phi$.
Therefore, an analogous construction for the R\"ontgen term of the UDW model would read
\begin{align}
  \hat H_{\text{R\"o}}= - \lambda\int_{\mathbb{R}^3} \dd^3 \bm R\left\{\frac{\hat{\bm P}}{2 M},\hat{\bm x}\times\grad_{\bm R}\hat{\phi}(\bm R)\right\}_+\dyad{\bm R}{\bm R}.
\end{align}
 Notice that, same as in the full non-relativistic light-matter interaction this term couples the internal degrees of freedom of the atom simultaneously with both the COM degrees of freedom and the field. This coupling cannot be expected to be any less significant for relativistic studies where the relativistic corrections induce additional couplings between all these degrees of freedom. Adding to the usual monopole coupling a term analogous to the R\"ontgen dynamics is thus necessary if one wants to mimic atom-light interactions, and if the COM is treated as a quantum degree of freedom.

Modifications to the UDW model beyond the correspondence with the non-relativistic multipolar Hamiltonian can be thought of along several routes. A leading order correction could come from the analogy  with the sub-leading Darwin Hamiltonian from Sec.~\ref{sec rel cor}. This however cannot provide a fully relativistic treatment, and so a non-perturbative approach is preferable to account for ultra-relativistic regimes. Secondly, so far the spin degrees of freedom have been neglected due to their sub-leading order effects. However, in relativistic regimes a coupling of those spin degrees of freedom with the other atomic and field degrees of freedom should be expected to become significant. A scalar construction inspired by the Breit Hamiltonian~\cite{standardgamma} may be a possible avenue of future work. 

\section{Conclusion}

In this study, we gave an account of the different levels of light-atom interaction models that are commonly encountered in the literature, in particular the RQI community. Our aim was to bring focus on a) the gauge issues that appear when considering the light-matter interaction when the center of mass of the atom is not classical and b) the often neglected dynamics arising from the center-of-mass degrees of freedom that are still leading order and important when considering quantum COM dynamics, such as R\"ontgen terms.

We reviewed, with a focus on the models typically used in RQI, the Hamiltonian formulation of the interaction of an atom with the electromagnetic field. We reviewed in detail the different ingredients of the derivation of the multipolar Hamiltonian of an atom interacting with quantized light at the dipole level in the non-relativistic COM motion approximation. Through this, we gave an account of the subtleties of the transformations between different Hamiltonian formulations for a two-particle atom. We discussed  the origin of the different terms in the Hamiltonian in the preferable, and physically motivated set of canonical variables for which we can understand the interaction as a hydrogen-like atom perturbed by a dynamical electromagnetic field. We discussed how relativistic COM motion needs to incorporate extra couplings between internal and external degrees of freedom of the atom when the COM is quantum. In the non-relativistic regime we discussed the importance of the R\"ontgen term for proper atomic dynamics, in particular at the example of transition rates and for a delocalized COM. We also discussed how the R\"ontgen cannot be cancelled by a choice of reference frame for a quantized center of mass.

We further showed that one can nonetheless consider relativistic atomic trajectories in simple ways if one is willing to neglect some of the atomic dynamics and treat the COM as classical. This allows to use a simple effective dipolar interaction model, which - in contrast to the non-relativistic multipolar Hamiltonian which is manifest Galilei invariant- satisfies a Lorentz-covariant prescription. Lastly, in the context of scalarized field-matter models, we suggested to modify the Unruh-DeWitt model if we want to account for the COM dynamics that comes through the R\"ontgen term and acts at the same order in all small parameters as the dipole term.

The discussions on this manuscript are intended to provide a closer and more detailed link between the simple particle detector models (ubiquitously employed in quantum field theory in curved spacetimes as well as in relativistic quantum information) and atomic physics, pointing out the subtleties regarding gauge transformations and choice of physical variables,  and identifying to what extent and in what regimes scalar models  capture essential features of the light-matter interaction. These notes pave the way to further studies of how relativistic motion of delocalized center-of-mass atoms influence typical protocols of relativistic quantum information, and will undoubtedly be studied elsewhere.

\section{Acknowledgements}				

The authors would like to thank T. Rick Perche for helpful discussions. This project is partially supported by the NSERC Discovery program. E. M-M acknowledges funding from his Ontario Early Researcher Award.

\appendix
\begin{widetext}

\section{Comparison of PZW and Dirac-Heisenberg transformation}\label{pzw check}
In this section we prove that the Dirac-Heisenberg transformation from eq.~\eqref{gauge 2} is identical to the PZW transformation of Eq.~\eqref{pzw gauge} to all orders. To do so, we expand the field around the position of the COM coordinate $\bm R$ and perform the integral over $\lambda$. From Eq.~\eqref{gauge 2} we get 
\begin{align}
   \hat\Lambda=&\int_{\mathbb{R}^3} \dd^3 \bm  R\int_{\mathbb{R}^3} \dd^3 \bm  r \dyad{\bm R}{\bm R}\otimes\dyad{\bm r}{\bm r}\int_{0}^{1} \dd \lambda\sum_{n=0}^\infty \frac{1}{n!}\bm r^n \cdot \bm\nabla^n_{\bm R}(\bm r\cdot\hat{\bm A}(t,\bm R)) \left(\lambda-\frac{m_e}{M}\right)^n\nonumber\\
   =&\int_{\mathbb{R}^3} \dd^3 \bm  R\int_{\mathbb{R}^3} \dd^3 \bm  r \dyad{\bm R}{\bm R}\otimes\dyad{\bm r}{\bm r} \sum_{n=0}^\infty \frac{\bm r^n \cdot \bm\nabla^n_{\bm R}(\bm r\cdot\hat{\bm A}(t,\bm R))}{(n+1)!}\left[\left(\frac{m_p}{M}\right)^{n+1}+(-1)^n\left(\frac{m_e}{M}\right)^{n+1}\right].\label{dirac check}
\end{align}
We can see that truncating after the first two terms yields $\hat\Lambda^{(1)}$, i.e. Eq.~\eqref{gauge 1}. Now we compare to the PZW   transformation:
\begin{align}
     \hat\Lambda^{\text{PZW}}=&\sum_{i=e,p} \frac{e_i}{|e|} (\hat{\bm r}^i-\hat{\bm R}) \cdot \int_{0}^{1} \dd \lambda\,  \hat{\bm{A}}\left(t,\hat{\bm R}+\lambda\big(\hat{\bm r}^i- \hat{\bm R}\big)\right)\nonumber\\
     =&\int_{\mathbb{R}^3} \dd^3 \bm  R\int_{\mathbb{R}^3} \dd^3 \bm  r \dyad{\bm R}{\bm R}\otimes\dyad{\bm r}{\bm r}\bm r \cdot \int_{0}^{1} \dd \lambda  \left[\frac{m_p}{M}\hat{\bm{A}}\left(t,\bm R+\lambda\frac{m_p}{M}\bm r\right)+\frac{m_e}{M}\hat{\bm{A}}\left(t,\bm R-\lambda\frac{m_e}{M}\bm r\right) \right]\nonumber\\
     =&\int_{\mathbb{R}^3} \dd^3 \bm  R\int_{\mathbb{R}^3} \dd^3 \bm  r \dyad{\bm R}{\bm R}\otimes\dyad{\bm r}{\bm r} \int_{0}^{1} \dd \lambda\sum_{n=0}^\infty \frac{ \lambda^n}{n!}\bm r^n \cdot \bm\nabla^n_{\bm R}(\bm r\cdot\hat{\bm A}(t,\bm R))\left[\left(\frac{m_p}{M}\right)^{n+1}+(-1)^n\left(\frac{m_e}{M}\right)^{n+1}\right]\nonumber\\
     =&\int_{\mathbb{R}^3} \dd^3 \bm  R\int_{\mathbb{R}^3} \dd^3 \bm  r \dyad{\bm R}{\bm R}\otimes\dyad{\bm r}{\bm r}\sum_{n=0}^\infty \frac{\bm r^n \cdot \bm\nabla^n_{\bm R}(\bm r\cdot\hat{\bm A}(t,\bm R)) }{(n+1)!}\left[\left(\frac{m_p}{M}\right)^{n+1}+(-1)^n\left(\frac{m_e}{M}\right)^{n+1}\right],\label{pzw check eq}
\end{align}
where for the second line we used
\begin{align}
    \hat{\bm r}_e=\hat{\bm R} +\frac{m_p}{M}\hat{\bm r},\quad \hat{\bm r}_p=\hat{\bm R}-\frac{m_e}{M}\hat{\bm r}.
\end{align}
We see that the PZW transformation of \eqref{pzw check eq} and the Dirac-Heisenberg transformation of \eqref{dirac check} are identical to all orders, and that, therefore, we will obtain the multipolar Hamiltonian after canonical transformation.

\section{Commutator computations}\label{commutators}
Through derivatives acting on 
\begin{align}
	    \left[ 	\hat{A}^i(t,\bm x),	\hat{A}^j(t,\bm x')\right]=&\int_{\mathbb{R}^3} \!\frac{\text{d}^3k}{(2 \pi)^{3}} \frac{\hbar }{2 \epsilon_0 c |\bm k|}    (\delta^{ij} -  e^i_{\bm k}e^j_{\bm k}) \left( e^{\ii  \bm{k} \cdot (\bm{x}-\bm{x}')}-e^{-\ii  \bm{k} \cdot (\bm{x}-\bm{x}')} \right), \label{comm 1 app}
	\end{align}
we find the  commutators
	\begin{align}
	    \left[ 	\hat{A}^i(t,\bm x),	\partial_l \hat{A}^j(t,\bm x')\right]=&\int_{\mathbb{R}^3} \!\frac{\text{d}^3k}{(2 \pi)^{3}} \frac{-\ii\hbar }{2 \epsilon_0 c |\bm k|} k_l   (\delta^{ij} -  e^i_{\bm k}e^j_{\bm k})\left( e^{\ii  \bm{k} \cdot (\bm{x}-\bm{x}')}+e^{-\ii  \bm{k} \cdot (\bm{x}-\bm{x}')} \right)\xrightarrow{\bm x=\bm x'}0,  \label{comm 2} \\
	    \left[ \partial_l\hat{A}^i(t,\bm x),	\partial_m \hat{A}^j(t,\bm x')\right]=&\int_{\mathbb{R}^3} \!\frac{\text{d}^3k}{(2 \pi)^{3}} \frac{\hbar }{2 \epsilon_0 c |\bm k|}k_m k_l  (\delta^{ij} -  e^i_{\bm k}e^j_{\bm k}) \left( e^{\ii  \bm{k} \cdot (\bm{x}-\bm{x}')}-e^{-\ii  \bm{k} \cdot (\bm{x}-\bm{x}')} \right)\xrightarrow{\bm x=\bm x'}0, \label{comm 3}\\
	    \left[ 	\hat{A}^i(t,\bm x),	\partial_{t}\partial_l \hat{A}^j(t,\bm x')\right]=&\int_{\mathbb{R}^3} \!\frac{\text{d}^3k}{(2 \pi)^{3}} \frac{ \hbar }{2 \epsilon_0 }  k_l  (\delta^{ij} -  e^i_{\bm k}e^j_{\bm k}) \left( e^{\ii  \bm{k} \cdot (\bm{x}-\bm{x}')}-e^{-\ii  \bm{k} \cdot (\bm{x}-\bm{x}')} \right)\xrightarrow{\bm x=\bm x'}0,  \label{comm 5} \\
	      \left[ \partial_l	\hat{A}^i(t,\bm x),	\partial_{t} \hat{A}^j(t,\bm x')\right]=&\int_{\mathbb{R}^3} \!\frac{\text{d}^3k}{(2 \pi)^{3}} \frac{ \hbar }{2 \epsilon_0 }  (-k_l)  (\delta^{ij} -  e^i_{\bm k}e^j_{\bm k}) \left( e^{\ii  \bm{k} \cdot (\bm{x}-\bm{x}')}-e^{-\ii  \bm{k} \cdot (\bm{x}-\bm{x}')} \right)\xrightarrow{\bm x=\bm x'}0,   \label{comm 6}\\
	       \left[ 	\hat{A}^i(t,\bm x),	\partial_{t} \hat{A}^j(t,\bm x')\right]=&\int_{\mathbb{R}^3} \!\frac{\text{d}^3k}{(2 \pi)^{3}} \frac{\ii \hbar }{2 \epsilon_0 }    (\delta^{ij} -  e^i_{\bm k}e^j_{\bm k}) \left( e^{\ii  \bm{k} \cdot (\bm{x}-\bm{x}')}+e^{-\ii  \bm{k} \cdot (\bm{x}-\bm{x}')} \right)=\frac{\ii \hbar}{\epsilon_0} \delta^{i j, (\text{tr})}(\bm x-\bm x'),  \label{comm 4 app} \\
	    \left[ \partial_l	\hat{A}^i(t,\bm x),	\partial_{t}\partial_m \hat{A}^j(t,\bm x')\right]=&\int_{\mathbb{R}^3} \!\frac{\text{d}^3k}{(2 \pi)^{3}} \frac{\ii \hbar }{2 \epsilon_0 }  k_l k_m  (\delta^{ij} -  e^i_{\bm k}e^j_{\bm k}) \left( e^{\ii  \bm{k} \cdot (\bm{x}-\bm{x}')}+e^{-\ii  \bm{k} \cdot (\bm{x}-\bm{x}')} \right)=\frac{\ii \hbar}{\epsilon_0} \pdv{\delta^{i j, (\text{tr})}(\bm x-\bm x')}{x^l}{x'^m},  \label{comm 7 app}
	\end{align}
	note that $\partial_{t}\hat{A}^j=-\hat{E}^j$, and $\delta^{i j, (\text{tr})}(\bm x)=\frac{1}{(2\pi)^3}\int_{\mathbb{R}^3} \dd^3 \bm  k (\delta^{ij} -  e^j_{\bm k}e^j_{\bm k}) e^{\ii \bm k\cdot \bm x}$ is the transverse delta function.
We see that in the coincidence limit $\bm x=\bm x'$, Eq.~\eqref{comm 1 app} -- \eqref{comm 6} vanish. Eq.~\eqref{comm 2} vanishes due to parity: there is always an odd number of powers of $k_i$ being integrated over the whole momentum space. The other commutators are zero due to the cancellations of the plane waves. As all spatial commutators (without any time derivative being involved) vanish, this implies that the vector potential is left invariant under the unitary operator generated by Eq.~\eqref{gauge 1}. 	As we work in the dipole approximation, no higher derivatives will be needed  to compute  $\hat{\tilde H}^{(1)}$  in Eq.~\eqref{new H}. 

\section{Relations between the different electromagnetic Wightman tensors}\label{app-wightmann}
The Wightman tensors can be derived in integral form by virtue of $\matrixel{0}{\hat{a}_{\bm k, s} \hat{a}^\dagger_{\bm k', s'}}{0}=\delta_{s,s'} \delta(\bm k - \bm k')$, and the completeness relations of the polarization vectors
\begin{align}
   \sum_{s=1}^2   \epsilon_{\bm k, s}^i   \epsilon_{\bm k, s}^j = \delta^{ij} -  e^j_{\bm k}e^j_{\bm k}=\mathcal{F}(\delta^{i j, (\text{tr})}), 
\end{align}
where $\mathcal{F}$ represents the Fourier transform from $\bm x$ to $\bm k$.
It turns out that knowing two Wightman tensors suffices to characterize all remaining Wightman tensors of the electromagnetic field strength tensor. Starting with the purely electric field case, we get

\begin{align}
   W^{ij}_E[t,t';\bm x,\bm x']&= \bra 0 \hat{E}^i(t,\bm x) \hat{E}^j(t',\bm x') \ket 0\nonumber\\
  &=\sum_{s,s'=1}^2\frac{\hbar c}{2(2 \pi)^{3}\epsilon_0} \int_{\mathbb{R}^3}  \dd^3 \bm  k\int_{\mathbb{R}^3}  \dd^3 \bm  k' \sqrt{|\bm k||\bm k'|} \epsilon_{\bm k, s}^i   \epsilon_{\bm k, s'}^j \matrixel{0}{\hat{a}_{\bm k, s}\hat{a}^\dagger_{\bm k', s'}}{0}e^{-\ii c(|\bm k|t-|\bm k'|t')} e^{\ii (\bm k \cdot \bm x - \bm k'\cdot\bm x')} \nonumber\\
   &=\frac{\hbar c}{2\epsilon_0} \int_{\mathbb{R}^3}  \frac{\dd^3 \bm  k}{(2 \pi)^{3}} |\bm k| e^{-\ii c|\bm k| (t-t')} e^{\ii \bm k \cdot (\bm x - \bm x')} \left( \delta^{ij} -  e^i_{\bm k} e^j_{\bm k}\right).
   \end{align}

For the purely magnetic Wightman function we make use of (assuming a right-handed orthonormal basis)
\begin{align}
    \bm e_{\bm{k}}\cross\bm \epsilon_{\bm k, s}= \begin{cases} 
      \bm \epsilon_{\bm k, 2}, &  s=1 \\
     -\bm \epsilon_{\bm k,1}, & s=2 \\
   \end{cases}\quad\Rightarrow\quad\sum_{s=1}^2(\bm e_{\bm{k}}\cross\bm \epsilon_{\bm k, s})^i(\bm e_{\bm{k}}\cross\bm \epsilon_{\bm k, s})^j=  \sum_{s=1}^2   \epsilon_{\bm k, s}^i   \epsilon_{\bm k, s}^j.
\end{align}
It follows then simply
\begin{align}
   &W^{ij}_B[t,t';\bm x,\bm x']=\bra 0 \hat{B}^i(t,\bm x) \hat{B}^j(t',\bm x') \ket 0=\frac{W^{ij}_E[t,t';\bm x,\bm x']}{c^2}.
   \end{align}
   For the remaining Wightman functions we need
   \begin{align}
    \sum_{s=1}^2(\bm e_{\bm{k}}\cross\bm \epsilon_{\bm k, s})^i \epsilon_{\bm k, s}^j&=\epsilon_{\bm k, 2}^i \epsilon_{\bm k, 1}^j -\epsilon_{\bm k, 1}^i \epsilon_{\bm k, 2}^j =-\epsilon^{i j k} (e_{\bm{k}})_k.
   \end{align}
   Hence, we find
   \begin{align}
  & W^{ij}_{BE}[t,t';\bm x,\bm x']=\bra 0 \hat{B}^i(t,\bm x) \hat{E}^j(t',\bm x') \ket 0=-\frac{\hbar }{2\epsilon_0} \int_{\mathbb{R}^3}  \frac{\dd^3 \bm  k}{(2 \pi)^{3}} |\bm k| e^{-\ii c|\bm k| (t-t')} e^{\ii \bm k \cdot (\bm x - \bm x')}     \epsilon^{ijk} (e_{\bm k})_k=W^{ji}_{EB}[t,t';\bm x,\bm x'].
\end{align}

	\end{widetext}	
	
		\bibliography{myref}
\end{document}